\documentclass[3p,times]{article}
\usepackage[margin=1in]{geometry}

\usepackage{amsmath,amsfonts,amssymb}
\usepackage{bm}
\usepackage{calc}

\usepackage{lineno}

\usepackage[figuresright]{rotating}

\usepackage{subcaption}
\usepackage{authblk}
\usepackage[numbers,sort&compress]{natbib}

\begin{document}

\title{An Extension of the Localized Artificial Diffusivity Method for Immiscible and High Density Ratio Flows}
\date{March 14, 2024}

\author[1]{Steven R. Brill}
\author[1]{Britton J. Olson}
\author[2]{Guillaume T. Bokman}

\renewcommand\Affilfont{\fontsize{9}{10.8}\itshape}
\affil[1]{Lawrence Livermore National Laboratory}
\affil[2]{ETH Zurich}

\maketitle

\begin{abstract}
The localized artificial diffusivity (LAD) method is widely regarded as the preferred multi-material regularization scheme for the compact finite difference method, because it is conservative, easy to implement, and generally robust for a wide range of multi-material problems. However, traditional LAD methods face significant challenges when applied to flows with large density ratios and when maintaining thermodynamic equilibrium across material interfaces. These limitations arise from the formulation of the artificial diffusivity flux and the reliance on enthalpy diffusion for interface regularization. Additionally, traditional LAD methods struggle to ensure stability under large density ratio conditions, fail to maintain a finite interface thickness, and are therefore unsuitable for modeling immiscible interfaces. In this work, we discuss the origins of these issues in traditional LAD methods and propose modifications which enable the simulation of large density ratio and immiscible flows. The proposed method targets the artificial diffusion fluxes at gradients and ringing in the volume fraction, rather than the mass fraction in traditional methods, to consistently regularize large density ratio interfaces. Furthermore, the proposed method introduces an artificial bulk density diffusion term to enforce equilibrium conditions across interfaces. To address the challenge of modeling immiscible flows, a conservative diffuse interface term is incorporated into the formulation to ensure a finite interface thickness. Specific consideration is taken in the design of the method to ensure that these crucial properties are maintained for $N$-material flows. The effectiveness of the proposed method is demonstrated through a series of canonical test cases, and its accuracy is validated by comparison with experimental data on micro-bubble collapse in water. These results highlight the method's robustness and its ability to overcome the limitations of traditional LAD approaches.
\end{abstract}

\section{Introduction}\label{sec:intro}
Multi-material flows occur in a wide variety of applications such as combustion~\cite{kuo2012fundamentals}, inertial confinement fusion~\cite{craxton2015direct}, targeted drug delivery~\cite{bokman2024impulse}, and additive manufacturing~\cite{zhu2021mixed}.  These problems are often simulated with a single-fluid approach where a single set of governing equations treats all of the materials as separate components of a single fluid. Each component has its own material properties, equations of state, and potentially physical terms, such as material strength for solids. Accurately capturing the material interfaces introduces a number of computational challenges such as: accurately capturing large jumps in material properties, location of interfaces, and maintaining thermodynamic equilibrium at interfaces~\cite{saurel2018diffuse}. 

Simulations of multi-material flows are often performed on Eulerian grids, Lagrangian frames, or with arbitrary Lagrangian-Eulerian (ALE) methods. Lagrangian and ALE methods have shown successes for capturing complex multi-material mixing, but tend to be diffusive and suffer from poorly conditioned cells and mesh tangling~\cite{darlington2001study,anderson2018high}. In this work, we use Eulerian grids because they allow for the use of high-order finite difference methods, which can capture the small-scale features of turbulence without over-dissipation~\cite{lele1992compact}. In the context of Eulerian methods, there are two main multi-material approaches: interface tracking and interface detection. Interface tracking involves explicitly evolving the location of the interface in methods such as the marker-in-cell method~\cite{harlow1965numerical} and front-tracking methods~\cite{unverdi1992front}. These methods accurately capture interface shapes, but require complex treatments at the interface.  Interface detection methods compute the location of the interface using the flow variables within the simulation, which allows for better capturing of topology changes and interface creation. Within interface detection methods, there are sharp and diffuse interface methods. Sharp interface methods capture the material interface as a single-point jump, which often requires special treatment to maintain numerical stability~\cite{tezduyar2006interface}. Additionally, methods such as the volume-of-fluid~\cite{benson2002volume} and level-set methods~\cite{aslam1996level} reconstruct the interface location for accuracy, but struggle to maintain exact conservation~\cite{olsson2005conservative}. Diffuse interface methods spread the material interface over multiple grid points, which avoids numerical instabilities, but introduces complexity with the treatment of points with multiple materials~\cite{saurel1999simple,saurel2018diffuse,allaire2002five,kapila2001two}. In this work, we are using a diffuse interface approach due to its simplicity of interface representation and conservation properties.

In this work, we are adopting a four-equation approach, where conservation of species mass, total momentum, and total energy are solved and points with multiple materials are assumed to instantaneously reach pressure and temperature equilibrium. In the literature, authors have studied the five~\cite{allaire2002five}, six~\cite{saurel1999simple}, and seven-equation models~\cite{kapila2001two} where pressure and temperature equilibrium are reached via relaxation terms and additional partial differential equations (PDEs) are solved for material volume fractions, internal energies, and velocities respectively. While these methods can be more expensive due to the additional PDEs, they have shown success for applications with large density ratios, disparate equations of state, and non-equilibrium effects~\cite{saurel1999simple,allaire2002five,kapila2001two}.

A wide range of numerical methods from low-order finite volume~\cite{jain2020conservative}, finite difference~\cite{aslani2018localized}, high-order discontinuous Galerkin~\cite{huang2023consistent}, and high-order finite difference methods~\cite{cook2009enthalpy} have shown success for simulating multi-material flows with the diffuse interface method. In this work, we are choosing to use high-order compact finite difference method~\cite{lele1992compact} due to excellent resolution of the small-scale features of turbulence. High-order methods introduce additional challenges in the context of multi-material flows because large differences in densities or material properties at interfaces can lead to Gibbs oscillations, which lead to unphysical values and code crashes~\cite{west2024high}. In order to address this issue, Cook introduced artificial-fluid large-eddy simulation (AFLES), where artificial fluid properties are added to the diffusive fluxes to stabilize unphysical oscillations while maintaining the excellent spectral properties of the underlying numerical method~\cite{cook2007artificial}. Then Cook added an enthalpy diffusion term to the multi-material treatment in AFLES to avoid violations of the entropy conditions and unphysical temperature spikes caused by the artificial species diffusion~\cite{cook2009enthalpy}. The artificial diffusivity method of AFLES was developed further for curvilinear grid and to improve its properties for turbulence and coined localized artificial diffusivity (LAD) methods~\cite{kawai2008localized,kawai2010assessment,kawai2011high,jain2023assessment,west2024high}. LAD methods have also been developed in the context of different underlying numerical schemes as well such as 2nd order finite difference methods and flux reconstruction~\cite{aslani2018localized,haga2019robust,jain2024stable,yu2013artificial,mirjalili2021localized}. One limitation of previous LAD methods is that they are unable to maintain equilibrium conditions at material interfaces without the adding additional PDEs and specifying specific equations of state~\cite{fujiwara2023fully}. Additionally, LAD methods struggle for large density ratios. 

Another weakness of LAD methods is that they continuously diffuse at the interface, so a finite interface thickness cannot be enforced. One common way to address this is through the phase-field method, where an additional PDE is solved to compute an additional flux term to enforce a constant interface thickness~\cite{huang2020consistent}. This method has seen success for a wide range of multi-material problems~\cite{mirjalili2021localized,mirjalili2022computational,mirjalili2024conservative,huang2023consistent,huang2024consistent,huang2025bound,white2025high,hwang2024robust}. The conservative diffuse interface (CDI) method is a particular choice of phase-field model where a conservative flux term is used to force multi-material interfaces toward a finite thickenss while maintaining conservation~\cite{chiu2011conservative,mirjalili2020conservative,jain2020conservative}. The accurate conservative diffuse interface (ACDI) method is another conservative phase-field model that formulates its sharpening flux with a signed distance function for improved accuracy \cite{jain2022accurate}. These methods have been successful at ensuring conservation while maintaining fixed interface thicknesses for immiscible interfaces within a variety of numerical methods~\cite{mirjalili2020conservative,jain2022accurate,mirjalili2021consistent,mirjalili2024conservative,mirjalili2023assessment,hwang2024robust,hatashita2025interface}. The CDI and ACDI methods have also been applied in the context of four-equation models where the volume fraction is computed from the conserved variables and is used as the phase-field variable without the addition of an extra PDE~\cite{jain2023assessment,collis2022assessment,collis2025thermodynamically}. 

The goal of this work is to improve upon existing LAD methods for the compact finite difference methods so that they can be applied to larger density ratio flows and immiscible flows. This development will be in the context of a four-equation model due to its simplicity, ease of enforcing continuity, and instantaneous enforcement of equilibrium. The LAD method will be modified and augmented with CDI terms in order to improve its stability and add the ability to capture immiscible interfaces while still maintaining uniform numerical treatment across the domain. Uniform treatment ensures the excellent spectral resolution of the compact finite difference method and parallel scalability are maintained. 

Sections~\ref{sec:eqns}-\ref{sec:num_meth} will detail the governing equations, numerical method, and the limitations of traditional methods for simulating large density ratios, maintaining equilibrium, and maintaining finite interface thicknesses. Section~\ref{sec:LAD} will explain our modifications to the traditional methods and how they address these limitations. Section~\ref{sec:results} will demonstrate these benefits on a range of test problems.

\section{Governing Equations}\label{sec:eqns}
The governing equations for multi-material flow in conservative Eulerian form are the equations of conservation of species mass, total momentum, and total energy:
\begin{align}
\frac{\partial \rho Y_i}{\partial t} &+ \nabla \cdot \left(\rho Y_i \bm{u}\right) = -\nabla\cdot \bm{J}_i, \label{eqn:specmass}\\
\frac{\partial \rho \bm{u}}{\partial t} &+ \nabla \cdot \left(\rho \bm{u} \otimes \bm{u} + p\bm{\delta}\right) = \nabla\cdot\bm{\tau} -\nabla \cdot \bm{F}, \label{eqn:mom} \\
\frac{\partial E}{\partial t} &+ \nabla \cdot\left[\left(E+p\right)\bm{u}\right] = \nabla\cdot\left(\bm{\tau}\cdot\bm{u} - \bm{q} - \bm{q}_d\right) - \nabla \cdot\bm{H}, \label{eqn:energy}
\end{align}
where $t$ is time, $\rho=\sum_{i=1}^{N_s}V_i \rho_i$ is the mixture density, $N_s$ is the number of species, $V_i$ is the volume fraction of material $i$, $\rho_i$ is the density of material $i$, $Y_i$ is the mass fraction of material $i$, $\bm{u}$ is the velocity, $\bm{J}_i$ is the species diffusion flux, $p$ is pressure, $\bm{\tau}$ is the stress tensor, $\bm{F}$ is the momentum consistency flux, $E$ is total energy, $\bm{q}$ is the thermal conductivity flux, $\bm{q}_d$ is the enthalpy flux between materials, and $\bm{H}$ is the energy consistency flux. The total energy is defined as 
\begin{align}
    E &= \frac{1}{2}\rho \bm{u}\cdot\bm{u} + \rho e, \\
    \rho e &= \sum_{i=1}^{N_s} \rho Y_i e_i,
\end{align}
where $e$ is the mixture specific internal energy and $e_i$ is the specific internal energy of material $i$. Mass fraction and volume fraction are related as $\rho Y_i = \rho_i V_i$. The viscous fluxes are defined as the sum of a physical viscous flux and an artificial flux for stabilization:
\begin{align}
    \bm{J}_i &= \bm{J}_{f,i} + \bm{J}_i^*, \\
    \bm{\tau} &= \bm{\tau}_{f} + \bm{\tau}^*, \\
    \bm{q} &= \bm{q}_{f} + \bm{q}^*, \\
    \bm{q}_d &= \bm{q}_{d,f} + \bm{q}^*_d, 
\end{align}
where the subscript $f$ is the physical fluid flux and the asterisks indicate the artificial fluxes. The physical fluxes for Fickian diffusion between species~\cite{bird2002transport}, viscous stress for Newtonian fluids, thermal conductivity, and enthalpy diffusion respectively are defined as~\cite{cook2009enthalpy}:
\begin{align}
    \bm{J}_{f,i} &= -D_{f,i} \rho \nabla Y_i + \rho Y_i \sum_{k=1}^{N_s} D_{f,k} \nabla Y_k, \label{eqn:Jphys} \\
    \bm{\tau}_f &= \mu_f(2\bm{S}) + \left(\beta_f -\frac{2}{3}\mu_f\right)(\nabla \cdot \bm{u})\bm{\delta}, \\
    \bm{q}_f &= -\kappa_f \nabla T , \label{eqn:qphys} \\
    \bm{q}_{d,f} &= \sum_{i=1}^{N_s} \bm{J}_{f,i} h_i , \label{eqn:qdphys}
\end{align}
where $D_{f,i}$ is the physical species diffusion coefficient, $\mu_f$ is the physical shear viscosity, $\beta_f$ is the physical bulk viscosity, $\bm{\delta}$ is the identity matrix, $\bm{S}$ is the symmetric strain rate tensor,
\begin{align}
\bm{S} = \frac{1}{2}[\nabla \bm{u} + (\nabla \bm{u})^T],
\end{align}
$\kappa_f$ is the physical thermal conductivity, and $h_i$ is the enthalpy of material $i$ as defined as
\begin{align}
    h_i = e_i + p_i/\rho_i.
\end{align}
The artificial fluxes will be defined in Section~\ref{sec:num_meth}. The separation of physical and numerical viscous fluxes allows the formulation to capture a wide range of scenarios. For example, in multi-material simulations of materials that diffuse into each other, $D_{f,i}$ is nonzero, but when that is not the case, $D_{f,i}$ is set to zero and only the artificial diffusion is used for stabilization.  The materials each have their own equation of state. In mixed grid points, the materials are assumed to the in pressure-temperature equilibrium, which is found using a Newton-iteration within each point as described in~\cite{cook2009enthalpy}.

\section{Numerical Method}\label{sec:num_meth}
We solve the equations above using the high-order finite difference code Miranda, developed at Lawrence Livermore National Laboratory. The Miranda hydrodynamics code has been used to study high Reynolds number turbulent, multi-species mixing~\cite{cook2007artificial,cook2009enthalpy,olson2014large,olson2007rayleigh,rehagen2017validation,morgan2018large}. Spatially, it uses a tenth-order compact finite difference scheme~\cite{lele1992compact}. Temporally, it uses a five-stage, fourth-order, low-storage, Runge-Kutta scheme~\cite{kennedy2000low}. Full numerical details are written in~\cite{cook2007artificial}. In order to dealias and improve numerical stability, an eighth-order compact filter is applied to the conserved variables to remove the top 10\% of wavenumbers. Because of the order of the method, Gibbs oscillations can be seen in under-resolved steep features, such as shocks and material interfaces. To address this issue, we use a localized artificial diffusivity (LAD) approach to regularize these steep features. The traditional method of formulating the LAD terms is using artificial fluid properties and formulating the diffusive fluxes in the same way as physical fluxes in Eqs.~\eqref{eqn:Jphys}-\eqref{eqn:qdphys}~\cite{cook2007artificial}. Hence, traditionally, the LAD terms have the form:
\begin{align}
    \bm{J}_{i}^* &= -D_{i}^* \rho \nabla Y_i + \rho Y_i \sum_{k=1}^{N_s} D_{k}^* \nabla Y_k, \label{eqn:Jafles} \\
    \bm{\tau}^* &= \mu^*(2\bm{S}) + \left(\beta^* -\frac{2}{3}\mu^*\right)(\nabla \cdot \bm{u})\bm{\delta}, \\
    \bm{q}^* &= -\kappa^* \nabla T , \label{eqn:qafles} \\
    \bm{q}^*_d &= \sum_{i=1}^{N_s} \bm{J}_{i}^* h_i , \label{eqn:qdafles}
\end{align}
where $D_{i}^*$ is the artificial diffusivity, $\mu^*$ is the artificial shear viscosity, $\beta^*$ is the artificial bulk viscosity, and $\kappa^*$ is the artificial thermal conductivity. With this formulation, $\sum_{i=1}^{N_s}\bm{J}_i^* =0$, so there is no movement of mass by the LAD terms. Hence, the consistency terms are zero: 
\begin{align}
\bm{F} &= 0, \\
\bm{H} &= 0.    
\end{align}
In order to localize the artificial diffusivity around under-resolved steep features, the artificial properties are designed to vanish in smooth regions and provide dissipation near discontinuities. In order to detect regions of under-resolution, the eighth-derivative operator is used:
\begin{align}
F(\phi) = \text{max}\left(\left|\frac{\partial^8 \phi}{\partial x^8}\Delta x^8\right|,\left|\frac{\partial^8 \phi}{\partial y^8}\Delta y^8\right|,\left|\frac{\partial^8 \phi}{\partial z^8}\Delta z^8\right|\right),
\end{align}
where $\phi$ is a scalar. For a vector, $\bm{\phi}$, 
\begin{align}
F(\bm{\phi}) = \text{max}(F(\phi_x),F(\phi_y),F(\phi_z)).
\end{align}
This introduces a $k^8$ wavenumber damping~\cite{cook2007artificial,morgan2018large}. The artificial diffusivities are
\begin{align}
    D^*_i &= \max \left[ C_{D,Y} \overline{F(Y_i)}\frac{\Delta^{2}}{\Delta t}, C_{O,Y}\overline{|Y_i|-1+|1-Y_i|}\frac{\Delta^{2}}{\Delta t} \label{eqn:Dafles} \right],\\
    \mu^* &= C_\mu \overline{F(\bm{\bm{u}})}\Delta, \\
    \beta^* &= C_\beta \overline{\rho F(\nabla\cdot \bm{u})}\Delta^{2}, \\
    \kappa^* &= C_\kappa \overline{\rho c_s^3F( 1/T)}\Delta^{2} ,
\end{align}
where $\Delta$ is the grid spacing, $\Delta t$ is the timestep size, $c_s$ is the speed of sound, $C_{D,Y}$, $C_{O,Y}$, $C_\mu$, $C_\beta$, and $C_\kappa$ are nondimensional constants chosen to control the strength of the artificial diffusion terms, and the overbar denotes a truncated-Gaussian filter:
\begin{align}
    \bar{f}(\bm{x}) &= \int_{-L}^L G(|\bm{x}-\bm{\xi};L)f(\bm{\xi})d^3\xi, \\
    G(\zeta;L) &= \frac{e^{-6\zeta^2/L^2}}{\int_{-L}^L e^{-6\zeta^2/L^2} d\zeta}, \quad L=4\Delta .
\end{align}
The Gaussian filter is used to remove cusps from the absolute values.

The two terms in Eq.~\eqref{eqn:Dafles} correspond to different uses for artificial diffusivity. The first term adds diffusivity to regions of ringing due to under-resolution. The second term is nonzero when the mass fraction is outside of 0 and 1, so diffusivity is added when the mass fraction is outside of its physical bounds. Typically, $C_{O,Y}$ is much larger than $C_{D,Y}$ so the artificial diffusivity quickly regularizes unphysical solutions. Some typical values of the coefficients are $C_{D,Y}=2e-4$, $C_{O,Y}=100$, $C_\mu=1e-4$, $C_\beta=7e-2$, and $C_\kappa=1e-3$. These values will be used for the remainder of the paper unless stated otherwise. 

\subsection{Bubble Advection Stability Test}\label{sec:pokeballtest}
\begin{figure}[]
    \centering
    \includegraphics[width=0.75\textwidth,trim={0 0 0 0cm}]{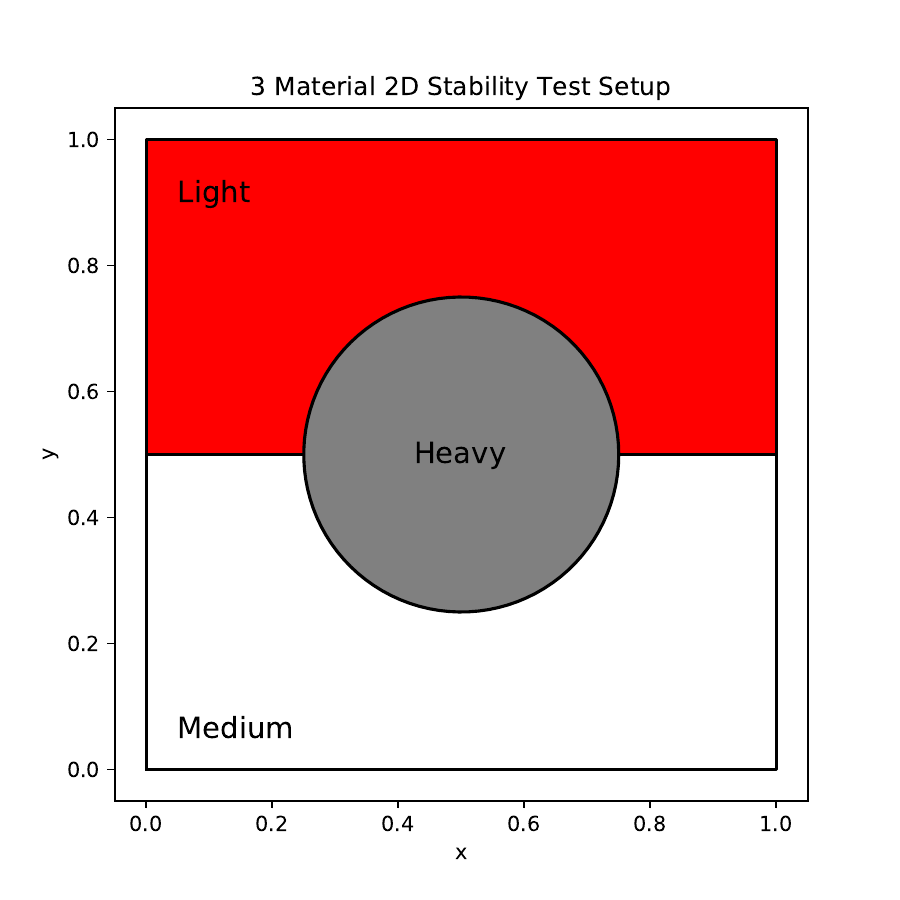}%
    \caption{Schematic of the initial setup for the bubble advection stability test.}
    \label{fig:pokeball}
\end{figure}

In order to study and demonstrate the limitations of the traditional multi-material methods for large density ratios, we introduce this test case. In this case, a high density ratio bubble is advected through a domain of two other materials. The nondimensional 2D domain is $[0,1]\times[0,1]$ and periodic in both directions. A circular bubble with radius $r=0.25$ of material $h$ with density $\rho_h = R$ is initialized in the center of the domain. Outside of the bubble, the domain is split into two sections. For $y>0.5$, material $l$ is defined with density $\rho_l = 1$. For $y<0.5$, material $m$ is defined with density $\rho_m = 10^{\log_{10}(R)/2}$. All three materials are $\gamma$-law gases with $\gamma = 1.4$ and $c_{v,i}=\frac{1}{\rho_i(\gamma-1)}$. The interfaces between the materials are defined smoothly using a hyperbolic tangent 
\begin{align}
V_i = \frac{1}{2}\left(1 + \tanh\left(\frac{r-r_0}{\frac{3 N_p\Delta}{16}}\right)\right), \label{eqn:srbtanh}
\end{align}
where $N_p$ is a chosen number of grid points in the interface thickness and $\Delta$ is the grid spacing. This form is chosen so that 99\% of the interface is contained within $N_p$ grid points centered around $r_0$. The number of grid points in the domain varies with $N_p$ such that the physical thickness of the interface is $0.05$ to prevent the two sides of the bubble from interacting. An image of the initial domain is shown in Fig~\ref{fig:pokeball}. The entire domain is initialized at equilibrium with $P=1$, $T=1$, $\bm{u}=[10,0]$. The Mach number in the heavy material is $M=u_1/\sqrt{\gamma P/R}=10\sqrt{R/\gamma}$, which gives a range of $M=26.73-8451$ for the density ratios of $R=10-10^6$ that are tested. The bubble is advected for $10$ periods. Whether the simulation ran to completion and the magnitude of the pressure oscillations at completion are reported. Simulations that did not run to completion fail because ringing at the material interface results in an imaginary speed of sound.  The study was also repeated with larger artificial diffusivities and lower timesteps. These changes show some improvement for the stability of the method for larger density ratios, but do not affect the overall trends. Hence, the study uses the default LAD coefficients from Sec.~\ref{sec:num_meth} and a CFL of 0.2 is used in order keep the timesteps comparable between different methods.

The test is set up so that it has a maximum density ratio of $R$ and an interface thickness of $N_p$ points. We would like to simulate as large of a density ratio as possible with as few points in the diffuse interface as possible. A three-material, 2D problem is used to demonstrate that the conclusions apply to $N_s>2$ cases and not just two-material cases. Testing with more than two materials is important for methods designed for an arbitrary number of materials because with two materials, the fact that $Y_l = 1 - Y_h$ introduces, often beneficial, symmetries between terms derived from a single mass fraction. For example, in the two material case, $\overline{F( Y_l)} = \overline{F( Y_h)}$ and $\overline{|Y_l|-1+|1-Y_l|}=\overline{|Y_h|-1+|1-Y_h|}$, so $D^*_l=D^*_h$. Such symmetries do not exist for more than two materials, so a case with a three-material interface test ensures robustness. The goal of the LAD term is to stabilize the simulation, so whether the simulation ran to the end time is the primary metric of success. The magnitude of the pressure oscillations shows the equilibrium errors introduced by the multi-material scheme.

\subsection{Limitations of LAD Formulation}
\label{sec:limitations}
In the context of multi-material problems, the traditional LAD method detailed in Section~\ref{sec:num_meth} has a number of limitations. First, the method cannot maintain equilibrium at an interface: if the materials at the interface are equal at constant pressure, temperature, and velocity, the method will introduce pressure, temperature, and velocity oscillations. At a phenomenological level, the method is designed to mimic physical processes as described in \cite{cook2009enthalpy}. Hence, it is designed such that $\sum_{i=1}^{N_s} J^*_i = 0$, so bulk density does not diffuse. In order to regularize the interface, the species diffusion flux diffuses mass fractions which result in enthalpy diffusion. Then, enthalpy diffusion introduces pressure gradients, which induce momentum which regularizes the interface. Hence, changes in temperature, pressure, and velocity are the mechanism by which the method regularizes the interface.

\begin{figure}[]
    \centering

    \includegraphics[width=0.5\textwidth,trim={0 0 0 0cm}]{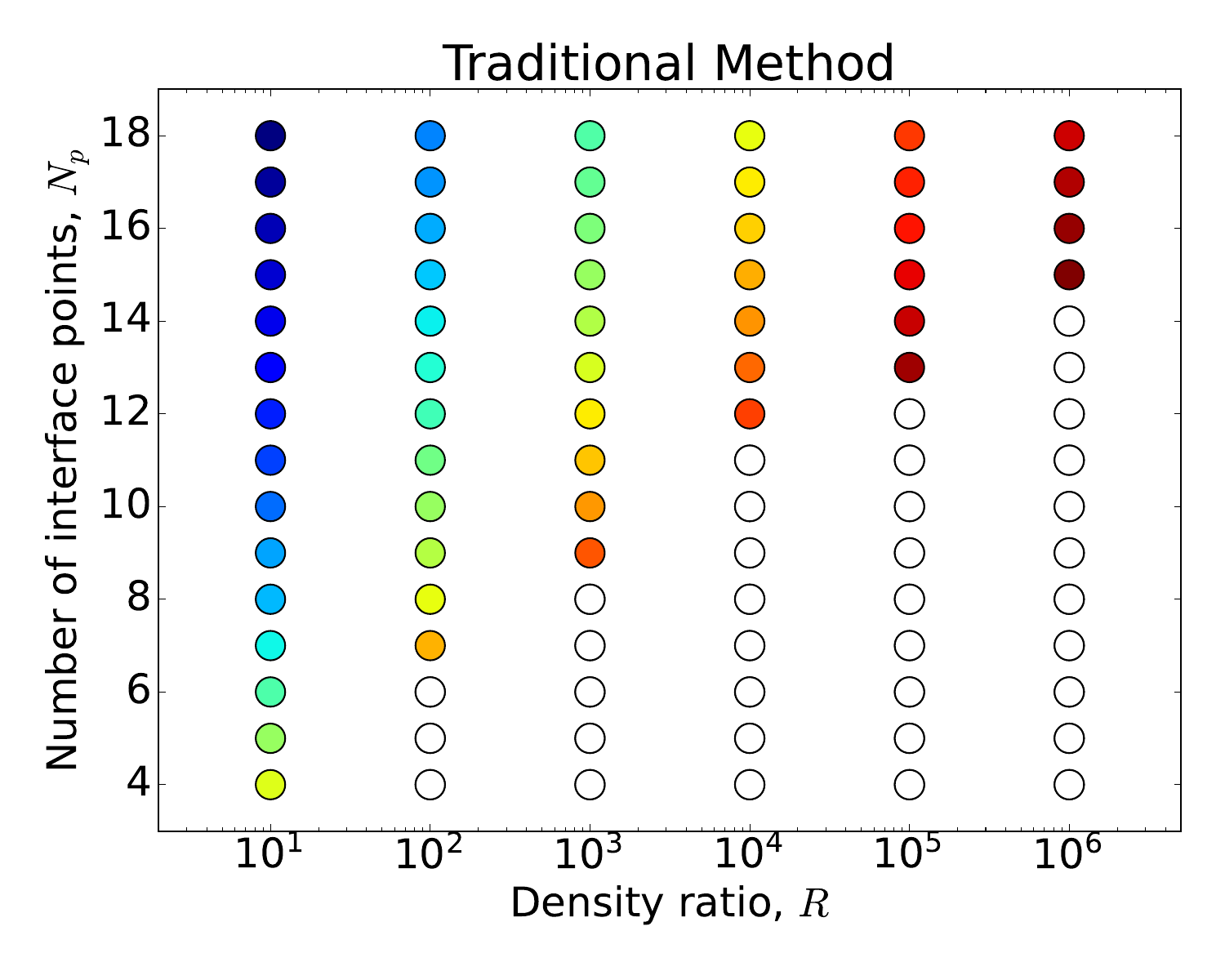}%
    \includegraphics[height=\heightof{\includegraphics[width=0.5\textwidth,trim={0 0 0 0cm}]{figures/stability_afles_new-eps-converted-to.pdf}}]{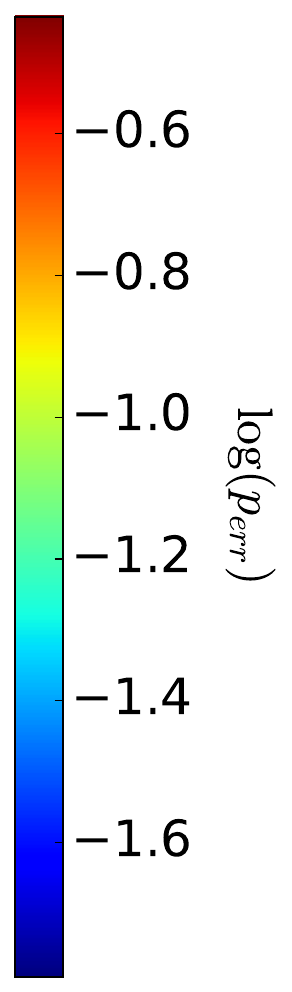}%
    \caption{Pressure oscillation magnitude and stability results for the 2D, three-material stability test for a range of number of points in the interface, $N_p$, and maximum density ratios between the materials, $R$. Empty circles indicate that the simulation went unstable before completion. The results presented are using the traditional method.}
    \label{fig:orig_stab}
\end{figure}

Additionally, for large density ratios, the method fails to regularize the interface and often goes unstable. To demonstrate this, we will study the advection of a high density bubble as detailed in Sec.~\ref{sec:pokeballtest}. Figure~\ref{fig:orig_stab} shows the results of this stability test for a range of $R$ and $N_p$. Empty markers indicate that the simulation went unstable before reaching completion. We see that for lower density ratios, the method remains stable for even $N_p=7$, while for $R>1000$, increasing numbers of points are needed to remain stable. Also, the pressure oscillations are at best around 1\% of the pressure and at worst greater than 25\%. This is because the method does not maintain equilibrium, as previously discussed. This performance leaves much to be desired because thick interfaces are necessary to stabilize large density ratios and the pressure oscillations are quite large, even in the best cases.

\begin{figure}[]
    \centering
    \includegraphics[width=0.75\textwidth,trim={0 0 0 0cm}]{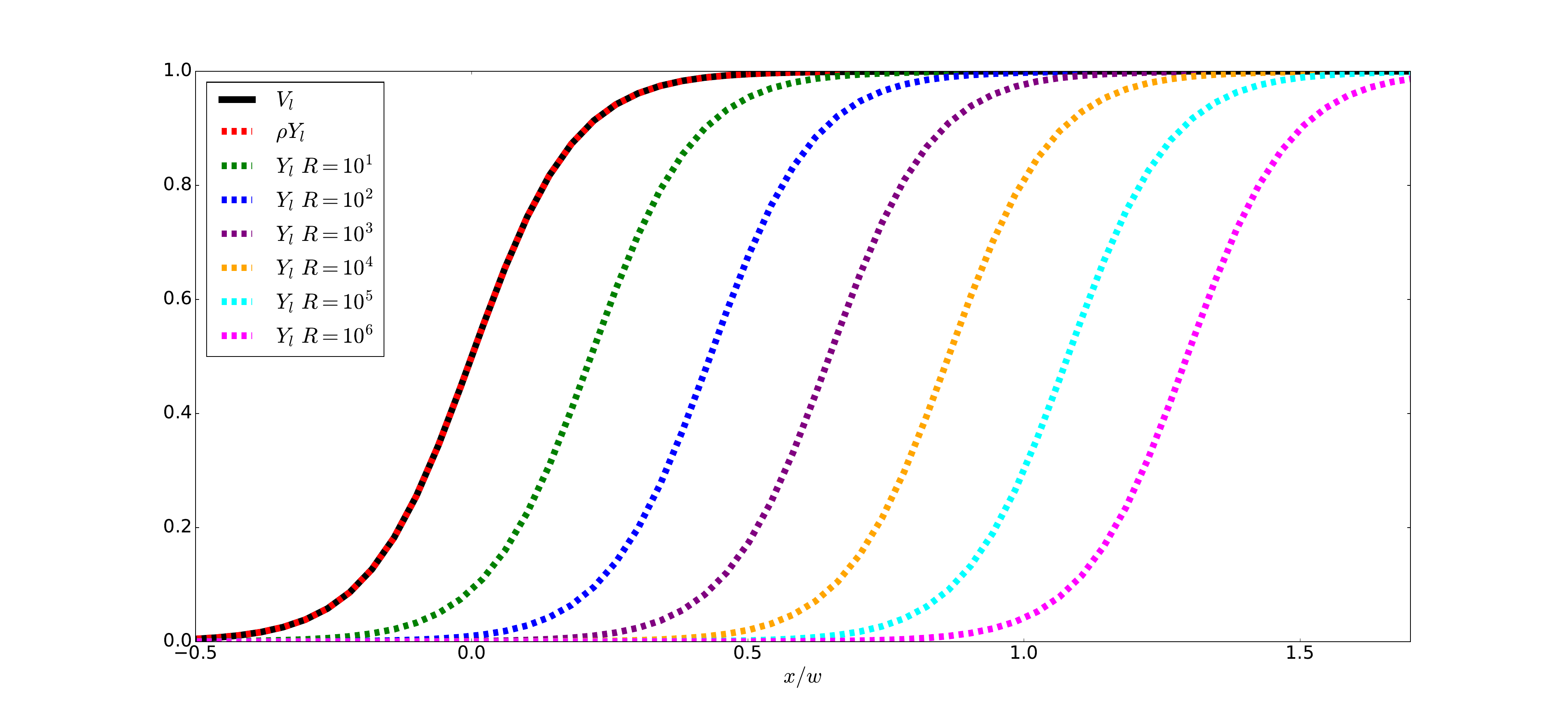}%
    \caption{Plot of the offset between volume fraction, $V_l$, and mass fraction, $Y_l$, for varying density ratios, $R$. The plotted interface has a hyperbolic tangent profile with width $w=0.1$. }
    \label{fig:offset}
\end{figure}

\begin{figure}[]
    \centering
    \includegraphics[width=0.5\textwidth,trim={0 0 0 0cm}]{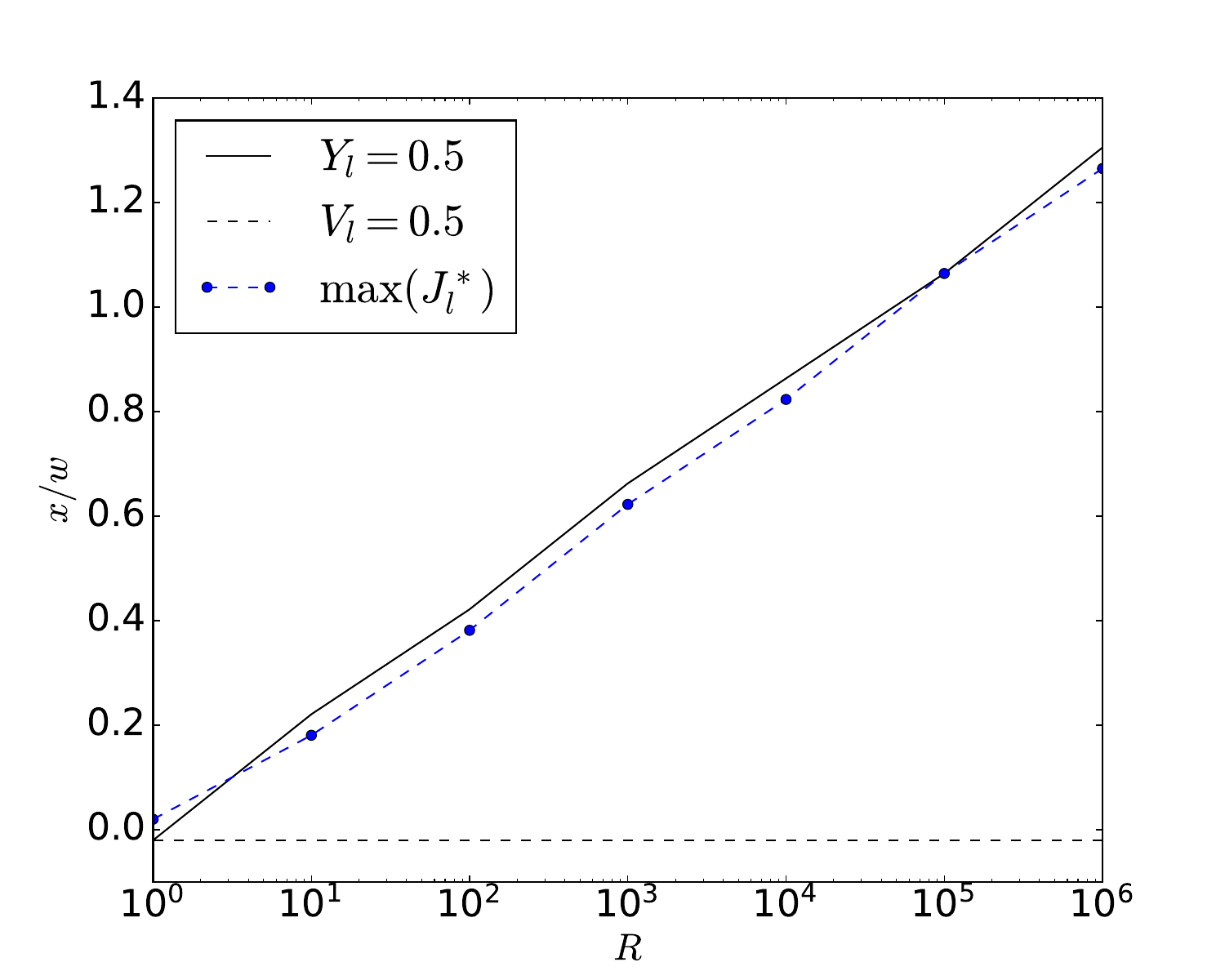}%
    \caption{Plot of the location of the maximum artificial species diffusion flux compared to the location of the interface for the mass fraction and volume fraction for varying density ratios. }
    \label{fig:flux_loc_orig}
\end{figure}

The stability test indicates that the existing scheme struggles for large density ratios. Now we will discuss the cause of this issue. In the four equation model presented, artificial species diffusion flux is used to regularize the species mass, as seen in Eq~\eqref{eqn:specmass}. Hence, the flux should address under-resolution in the species mass, which is also co-located with the volume fraction. As seen in Eqs.~\eqref{eqn:Jafles} and~\eqref{eqn:Dafles}, the diffusivity and flux target the mass fraction. In diffuse interface methods, because of the algebraic relationship between mass and volume fractions, the locations of the interface in the mass and volume fractions are not the same and are further separated as the mass fraction increases. For a 1D, 2 material interface, assuming the volume fraction interface is a hyperbolic tangent with width $w$, the offset between midpoints of the mass and volume fraction profiles, $x_{c,Y}$ and $x_{c,V}$ respectively, is
\begin{align}
    x_{c,Y} - x_{c,V} &= w \tanh^{-1}\left(-\frac{R-1}{R+1}\right) = w \tanh^{-1}\left( At \right),
\end{align}
where $At=\frac{\rho_l - \rho_h}{\rho_l + \rho_h}$ is the Atwood number. A full derivation is provided in~\ref{app:offset}. Figure~\ref{fig:offset} shows the species mass, volume fraction, and mass fraction for different density ratios. For lower density ratios, the species mass and mass fraction are reasonably well aligned, but as the density ratio increases, they become further offset. Hence, because the artificial diffusion fluxes are targeted at the mass fraction, as the density ratio increases, the fluxes become poorly localized, which leads to instabilities. Figure~\ref{fig:flux_loc_orig} shows the location of the maximum species diffusivity and the location of the interface as the density ratio increases, which demonstrates the poor localization.

Additionally, the method cannot capture immiscible interfaces. The regularization term only contains a diffusion term which will continuously diffuse the materials into one another resulting in a thick interface of mixed materials. While this assumption is appropriate for a range of problems involving gases that atomically mix, it does not allow for sharp interfaces to be maintained, like at the interfaces involving liquids and gases. Section~\ref{sec:LAD} will detail the development of an interface sharpening term that allows the presented method to model immiscible interfaces. 

\section{Localized Artificial Diffusivity Method}\label{sec:LAD}
In this section we will discuss the development of a new LAD method for an arbitrary number of materials that improves the stability of compact finite difference schemes for large density ratios, maintains equilibrium, and can simulate immiscible materials. First, we will list new terms and then we will discuss how the formulation accomplishes these goals. The new LAD flux and diffusivity formulations are:
\begin{align}
    \bm{J}^*_i &= -D^*_{\rho} \nabla(\rho Y_i) + Y_i \sum_{k=1}^{N_s} D^*_{\rho} \nabla(\rho Y_k) - Y_i D^*_\rho \nabla \rho = -D^*_{\rho} \nabla(\rho Y_i), \\
    D^*_{\rho} &= \max \left[D^*_{D,Y},D^*_{D,V}\right] + \max \left[D^*_{O,Y},D^*_{O,V}\right], \\
    D^*_{D,Y} &= \sum_{i=1}^{N_s} V_i \left[ C_{D,Y} \overline{|c_s F(Y_i)|}\Delta \right], \\
    D^*_{D,V} &= \sum_{i=1}^{N_s} V_i \left[ C_{D,V} \overline{|c_s F(V_i)|}\Delta \right], \\
    D^*_{O,Y} &= \max_{i=1,...,N_s} \left[ C_{O,Y}\overline{(|Y_i|-1+|1-Y_i|)c_s}\Delta \right], \\ 
    D^*_{O,V} &= \max_{i=1,...,N_s} \left[ C_{O,V}\overline{(|V_i|-1+|1-V_i|)c_s}\Delta \right], 
\end{align}
where $D^*_{\rho}$ is an artificial bulk density diffusivity and $C_{D,V}$ and $ C_{O,V}$ are additional coefficients, which are generally set equal to $C_{D,Y}$ and $C_{O,Y}$ respectively.  Unless otherwise stated, it is assumed that $C_{D,V}=2e-4$ and $C_{O,V}=100$. In this formulation there is diffusion of bulk density because $\sum_{i=1}^{N_s} \bm{J}_i^* = -D^*_{\rho} \nabla \rho$. Hence, consistency terms are added to the momentum and energy equations to account for the momentum and energy moved with the materials:
\begin{align}
    \bm{F}_i &= \left(\sum_{i=1}^{N_s} \bm{J}^*_i\right) \cdot \bm{u} , \label{eqn:momConsist}\\
    \bm{H} &= \left(\sum_{i=1}^{N_s} \bm{J}^*_i \right) \left(\frac{1}{2} \bm{u}\cdot \bm{u} \right), \label{eqn:eConsist} \\
    \bm{q}_d^* &= \sum_{i=1}^{N_s} \bm{J}^*_i e_i . \label{eqn:qdConsist}
\end{align}

\begin{figure}[]
    \centering
    \includegraphics[width=0.5\textwidth,trim={0 0 0 0cm}]{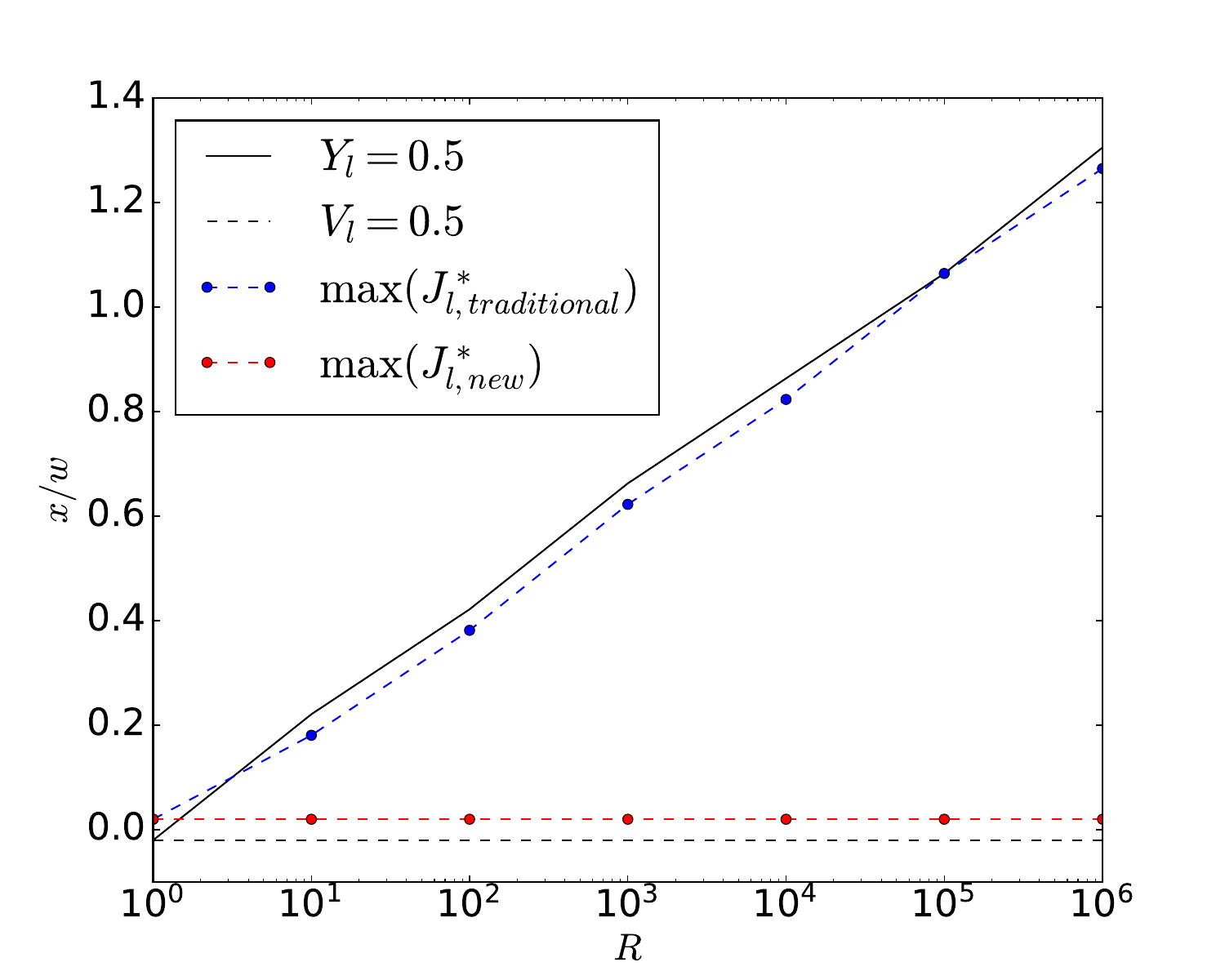}%
    \caption{Plot of the location of the maximum artificial species diffusion flux compared to the location of the interface for the mass fraction and volume fraction for varying density ratios for the new method compared to the traditional. }
    \label{fig:flux_loc_new}
\end{figure}

In order to regularize under-resolved interfaces, this method directly diffuses the species masses at the interface. Hence, $\bm{q}_d^*$ is only needed for consistency, not regularization, so it only contains internal energy and not enthalpy. The flux term uses gradients in $\rho Y_i$ so that it is localized around the interface in the transported quantity. In addition, to localize the diffusivity near the interface, even with large density ratios where there is an offset between mass fraction and species mass, diffusivity is targeted at ringing and out of bounds in both mass fraction and volume fraction. From numerical tests, even though the volume fraction based diffusivities are better aligned with the interface, keeping both mass and volume fraction terms improved stability for cases with large velocities and shocks. Figure~\ref{fig:flux_loc_new} shows the location of the new LAD term compared to the location of the interface. Unlike the previous formulation, the presented flux is aligned with the interface even as the density ratio increases.

A single diffusivity is used in all of the materials in order to prevent unphysical oscillations, which will be explained in detail in Sec.~\ref{sec:rhodiff}. In cases with two materials, the balance between the two mass fractions naturally created equal diffusivities. However, in the $N_s>2$ case, it is necessary to force the diffusivities to be equal. The overshoot diffusivities, $D^*_{D,Y}$ and $D^*_{D,V}$, are combined using the maximum so that the diffusivity addresses the worst overshoots. The ringing diffusivities, $D^*_{O,Y}$ and $D^*_{O,V}$, are volume fraction weighted to avoid large amounts of diffusivity being added when only trace amounts of a material are present. 

\subsection{Equilibrium}\label{sec:eqm}
Now we will show that the presented method maintains equilibrium conditions. Let there be an $N_s$ material, 1D interface at equilibrium with $P=P_0$, $T=T_0$, and $u=u_0$. At equilibrium, Eqs.~\eqref{eqn:specmass}-\eqref{eqn:energy} reduce to:
\begin{align}
    \frac{\partial \rho Y_i}{\partial t} &+ u_0 \nabla (\rho Y_i) = -\nabla \bm{J}^*_i, \\
    \frac{\partial \rho u}{\partial t} &+ u_0^2\nabla \rho = -u_0 \nabla \cdot \left(\sum_{i=1}^{N_s} \bm{J}^*_i\right),\\ 
    \frac{\partial E}{\partial t}  &+ u_0 \nabla E = -\nabla \cdot \bm{q}_d - \frac{1}{2} u_0^2 \nabla \cdot \left( \sum_{i=1}^{N_s} \bm{J}^*_i \right) .
\end{align}
Combining the species mass equations to the total density equation and rearranging the equations give:
\begin{align}
    \frac{\partial \rho}{\partial t} &= - u_0 \nabla \rho  - \nabla \left(\sum_{i=1}^{N_s} \bm{J}^*_i\right), \\
    \frac{\partial \rho Y_i}{\partial t} &= - u_0 \nabla (\rho Y_i) -\nabla \bm{J}^*_i, \\
    \frac{\partial \rho u}{\partial t}  &= - u_0^2\nabla \rho -u_0 \nabla \cdot \left(\sum_{i=1}^{N_s} \bm{J}^*_i\right),\\ 
    \frac{\partial E}{\partial t} &= -\frac{1}{2}u_0^3\nabla\rho - u_0 \nabla (\rho e) -\nabla \cdot \bm{q}_d - \frac{1}{2} u_0^2 \nabla \cdot \left( \sum_{i=1}^{N_s} \bm{J}^*_i\right) .
\end{align}
Considering an arbitrary equation of state of the form $\rho e = \rho e (p, \rho Y_i)$ gives a representation of the internal energy from the chain rule:
\begin{align}
    \frac{\partial \rho e}{\partial t} = \left(\frac{\partial \rho e}{\partial p}\right)_{\rho Y_i} \frac{\partial p}{\partial t} + \sum_{i=1}^{N_s} \left(\frac{\partial \rho e}{\partial \rho Y_i}\right)_{p,\rho Y_{j\neq i}} \frac{\partial \rho Y_i}{\partial t} . \label{eqn:drhoechain}
\end{align}
Expanding the internal energy to evaluate the derivative gives:
\begin{align}
    \rho e &= \sum_{i=1}^{N_s} \rho Y_i e_i, \\
    \left(\frac{\partial \rho e}{\partial \rho Y_i}\right)_{p,\rho Y_{j\neq i}} &= e_i .
\end{align}
To show equilibrium, it must be shown that the time derivatives of $u$, $p$, and $T$ are exactly zero. First considering velocity:
\begin{align}
    \frac{\partial u}{\partial t} &= \frac{\partial \left(\frac{\rho u}{\rho}\right)}{\partial t}, \\
    & = \frac{1}{\rho} \frac{\partial \rho u}{\partial t}  -\frac{u}{\rho} \frac{\partial \rho }{\partial t} \\ 
    & = \frac{1}{\rho} \left(- u_0^2\nabla \rho -u_0 \nabla \cdot \left(\sum_{i=1}^{N_s} \bm{J}^*\right) \right) - \frac{u_0}{\rho}\left( - u_0 \nabla \rho  - \nabla \left(\sum_{i=1}^{N_s} \bm{J}^*_i\right) \right), \\
    & = \left(\frac{u_0^2}{\rho} - \frac{u_0^2}{\rho}\right)\nabla \rho + \left(\frac{u_0}{\rho} - \frac{u_0}{\rho} \right)\nabla \left(\sum_{i=1}^{N_s} \bm{J}^*_i\right), \\
    & = 0.
\end{align}
Now for pressure we start by rearranging Eq.~\eqref{eqn:drhoechain}:
\begin{align}
    \frac{\partial p}{\partial t} &= \left(\frac{\partial \rho e}{\partial p}\right)_{\rho Y_i}^{-1} \left[ \frac{\partial \rho e}{\partial t} - \sum_{i=1}^{N_s} \left(\frac{\partial \rho e}{\partial \rho Y_i}\right)_{p,\rho Y_{j\neq i}} \frac{\partial \rho Y_i}{\partial t} \right], \\
    &= \left(\frac{\partial \rho e}{\partial p}\right)_{\rho Y_i}^{-1} \left[\frac{\partial E}{\partial t} - \frac{1}{2}u_0^2\frac{\partial \rho}{\partial t}  - \sum_{i=1}^{N_s} e_i \frac{\partial \rho Y_i}{\partial t} \right], \\
    & = \left(\frac{\partial \rho e}{\partial p}\right)_{\rho Y_i}^{-1} \left[-\frac{1}{2}u_0^3\nabla\rho - u_0 \nabla (\rho e) -\nabla \cdot \bm{q}_d - \frac{1}{2} u_0^2 \nabla \cdot \left( \sum_{i=1}^{N_s} \bm{J}^*_i\right) - \frac{1}{2}u_0^2\left(- u_0 \nabla \rho  - \nabla \left(\sum_{i=1}^{N_s} \bm{J}^*_i\right)\right) - \sum_{i=1}^{N_s} e_i \frac{\partial \rho Y_i}{\partial t} \right], \\
    & = \left(\frac{\partial \rho e}{\partial p}\right)_{\rho Y_i}^{-1} \left[- u_0 \nabla (\rho e)  -\nabla \cdot \bm{q}_d - \sum_{i=1}^{N_s} e_i \frac{\partial \rho Y_i}{\partial t}\right], \\
    & = \left(\frac{\partial \rho e}{\partial p}\right)_{\rho Y_i}^{-1} \left[- u_0 \nabla (\rho e)  -\nabla \cdot \bm{q}_d - \sum_{i=1}^{N_s} e_i \left(- u_0 \nabla (\rho Y_i) -\nabla \bm{J}^*_i \right) \right], \\
    &= \left(\frac{\partial \rho e}{\partial p}\right)_{\rho Y_i}^{-1} \left[- u_0 \nabla (\rho e)  -\nabla \cdot \bm{q}_d + u_0 \nabla \left( \sum_{i=1}^{N_s} \rho Y_i e_i \right) + \nabla\left(\sum_{i=1}^{N_s} \bm{J}^*_i e_i \right) \right], \\
    &= \left(\frac{\partial \rho e}{\partial p}\right)_{\rho Y_i}^{-1} \left[- u_0 \nabla (\rho e)  -\nabla \cdot \bm{q}_d + u_0 \nabla ( \rho e ) + \nabla\left(\sum_{i=1}^{N_s} \bm{J}^*_i e_i \right) \right], \\
    &= \left(\frac{\partial \rho e}{\partial p}\right)_{\rho Y_i}^{-1} \left[-\nabla \cdot \bm{q}_d + \nabla\left(\sum_{i=1}^{N_s} \bm{J}^*_i e_i \right) \right], \label{eqn:dpqd}\\
    &= \left(\frac{\partial \rho e}{\partial p}\right)_{\rho Y_i}^{-1} \left[-\nabla \cdot \left(\sum_{i=1}^{N_s} \bm{J}^*_i e_i \right) + \nabla\left(\sum_{i=1}^{N_s} \bm{J}^*_i e_i \right) \right], \\
     &= 0 .
\end{align}
As a note, for the traditional method, the derivation is exactly the same, except Eq.~\eqref{eqn:qdafles} is substituted into Eq.~\eqref{eqn:dpqd} leading to:
\begin{align}
 \frac{\partial p}{\partial t} &= \left(\frac{\partial \rho e}{\partial p}\right)_{\rho Y_i}^{-1} \left[-\nabla \cdot \left(\sum_{i=1}^{N_s} \bm{J}^*_i h_i \right) + \nabla\left(\sum_{i=1}^{N_s} \bm{J}^*_i e_i \right) \right], \\
 &= \left(\frac{\partial \rho e}{\partial p}\right)_{\rho Y_i}^{-1} \left[-\nabla \cdot \left(\sum_{i=1}^{N_s} \bm{J}^*_i \frac{P_0}{\rho_i} \right) \right],
\end{align}
which is not zero in general. The same analysis can be repeated to determine that $\frac{\partial T}{\partial t}=0$ at equilibrium for the presented method as well. 

\subsection{Sharpening}
In addition to improving the stability of interfaces with large density ratios, we also would like to model sharp interfaces, such as those occurring in flows with liquids and solids. The phase-field, CDI sharpening term from~\cite{mirjalili2024conservative} is formulated as a flux using the volume fraction as the phase-field variable following the approach of~\cite{jain2023assessment}. The resulting sharpening term is
\begin{align}
\bm{J}^*_{sharp,i} &= -\rho_i \Gamma \left[\epsilon \nabla V_i - \sum_{i\neq j}V_i V_j \hat{\bm{n}}_{ij}\right], \label{eqn:sharp_term} \\
\hat{\bm{n}}_{ij} &= \frac{\overline{\nabla V_{ij}}}{|\overline{\nabla V_{ij}}|}, \\
V_{ij} &= \frac{V_i}{V_i + V_j},
\end{align}
where $\Gamma$ is the velocity scale parameter, $\epsilon$ is the thickness parameter, $\hat{\bm{n}}_{ij}$ is the pairwise normal, and $V_{ij}$ is the pairwise volume fraction. The second term of Eq.~\eqref{eqn:sharp_term} sharpens the interface when two materials are present. Pairs of materials are considered here, so that $\sum_{i=1}^{N_s}\bm{J}^*_{sharp,i} = 0$ when the densities of all materials are equal. This prevents diffusion of bulk density from the presence of material interfaces in like materials. The first term of Eq.~\eqref{eqn:sharp_term} is an additional diffusion term to prevent over-sharpening. We set $\epsilon = O(\Delta)$, so that the interfaces are not sharpened smaller than the grid scale. $\Gamma$ is a parameter that sets the timescale at which the sharpening term acts. It is assumed that the advection timescale is dominant, so $\Gamma = O(\max(|\bm{u}|))$ to maintain sharp interfaces throughout the simulation. Unless stated otherwise, when sharpening is active, it is assumed that $\epsilon = \Delta$ and $\Gamma = \max(|\bm{u}|)$.

Including the sharpening term, the total species diffusive flux is:
\begin{align}
    \bm{J}_i^* = -D_\rho^* \nabla (\rho Y_i) -\rho_i \Gamma \left[\epsilon \nabla V_i - \sum_{i\neq j}V_i V_j \hat{\bm{n}}_{ij}\right] .
\end{align}
The consistency terms, Eqs.~\eqref{eqn:momConsist}-\eqref{eqn:qdConsist}, are applied with the total diffusive flux, so the equilibrium proof in Sec.~\ref{sec:eqm} is still valid with the sharpening term. The sharpening term is a flux term, so the method will remain conservative.

\subsection{Avoiding Unphysical Bulk Density Diffusion}\label{sec:rhodiff}
In the presented method, artificial bulk density diffusion is a mechanism by which interfaces are regularized. The diffusion term and the sharpening term are designed to ensure they do not cause bulk density change at interfaces between materials with equal densities. This can be shown as follows:

Assume the domain contains $N_s$ materials of densities $\rho_i = \rho_0$. Summing Eq.~\eqref{eqn:specmass} across all materials gives:
\begin{align}
    \frac{\partial \rho}{\partial t} + \nabla\cdot(\rho \bm{u}) &= -\nabla\cdot\left(\sum_{i=1}^{N_s}\bm{J}^*_i\right) \\
    \frac{\partial \rho}{\partial t} + \rho_0 \nabla\cdot(\bm{u}) &= -\nabla\cdot \left(\sum_{i=1}^{N_s}\left( -D_\rho^* \nabla (\rho Y_i) -\rho_i \Gamma \left[\epsilon \nabla V_i - \sum_{i\neq j}V_i V_j \hat{\bm{n}}_{ij}\right]\right) \right) \label{eqn:constrhosum}\\
    &= \nabla \cdot (D_\rho^*\nabla\rho_0) + \nabla\cdot\left(\rho_0\Gamma\left[\nabla\left(\sum_{i=1}^{N_s} V_i\right) - \sum_{i=1}^{N_s}\sum_{i\neq j}V_i V_j \hat{\bm{n}}_{ij}\right]\right) \\
    &=  \nabla \cdot (D_\rho^*\nabla\rho_0) \\
    &= 0.
\end{align}
Hence, the artificial diffusion terms do not cause a change in bulk density if all of the materials have the same density. This is the result of two main choices in the design of the method. First, the artificial species diffusivity for all materials are set to the same value, $D_\rho^*$. This choice allows the sum over the materials to enter into the gradient of the first term on the right-hand side of Eq.~\eqref{eqn:constrhosum} and ultimately go to zero. In Eq.~\eqref{eqn:Dafles}, the traditional method computes a separate material diffusivities based on ringing and overshoots of each material's mass fraction. For the case of two materials, $\overline{F( Y_l)} = \overline{F( Y_h)}$ and $\overline{|Y_l|-1+|1-Y_l|}=\overline{|Y_h|-1+|1-Y_h|}$, so with the traditional method, $D^*_l=D^*_h$, which satisfied the desired condition. However, for more than two materials, there are no such guarantees for the traditional diffusivity, so it could cause diffusion of density in a constant density interface. Hence, the choice of a common diffusivity across all materials is crucial to the formulation for more than two materials.

The second design choice is the use of the N-phase CDI formulation from Mirjalili \cite{mirjalili2024conservative} for the sharpening. Considering the materials pairwise while sharpening ensures that the sharpening terms sum to zero when summed across all materials. Without such a formulation, the density of constant density flows could diffuse from the sharpening term. Section~\ref{sec:shearflow} shows a three material, constant density case that demonstrates this strength of the presented method.

\subsection{Comparison with Artificial Mass Diffusion Methods}
In this section, we discuss the differences between the presented method and other LAD methods using artificial mass diffusion in the literature. The work of Terashima \cite{terashima2013consistent} uses mass diffusion for stabilization of multi-material interfaces without sharpening. To enforce interface equilibrium, they restrict their materials to calorically perfect gases and introduce a PDE to solve for the mixture ratio of specific heats. Aslani \cite{aslani2018localized} introduced a five-equation model for stiffened gases with artificial mass diffusion with no sharpening. They achieve interface equilibrium through a similar formulation of consistent fluxes as presented, but formulate the energy equation for the stiffened gas equation of state. The work of Jain \cite{jain2024stable} presents a five-equation model with a similar diffusion and sharpening formulation for two material flows. Instead of formulating the artificial diffusion fluxes in a consistent way, they use carefully constructed numerical fluxes to ensure interface equilibrium is preserved. The work of Collis \cite{collis2025thermodynamically} presents a four-equation model for multi-material and multi-phase flows using a CDI formulation for sharpening. This formulation uses similar consistent flux terms, but enthalpy is used in the energy consistency term, as opposed to internal energy in the presented formulation. Additionally, it is restricted to stiffened-gas equations of state and only tests two-material cases. The presented method expands upon the existing literature by ensuring interface equilibrium is maintained in a four-equation model for any number of materials without any assumptions on the equations of state. This is primarily achieved through the choices of a single artificial bulk density diffusivity, $D_{\rho}^*$, across all materials, careful derivation of the consistency requirements in Eqs.~\eqref{eqn:momConsist}-\eqref{eqn:qdConsist}, and the formulation of the sharpening term considering pairs of materials.

\section{Results}\label{sec:results}
\subsection{Stability Test}
\begin{figure}[]
    \centering
    \includegraphics[width=0.3\textwidth,trim={0 0 0 0cm}]{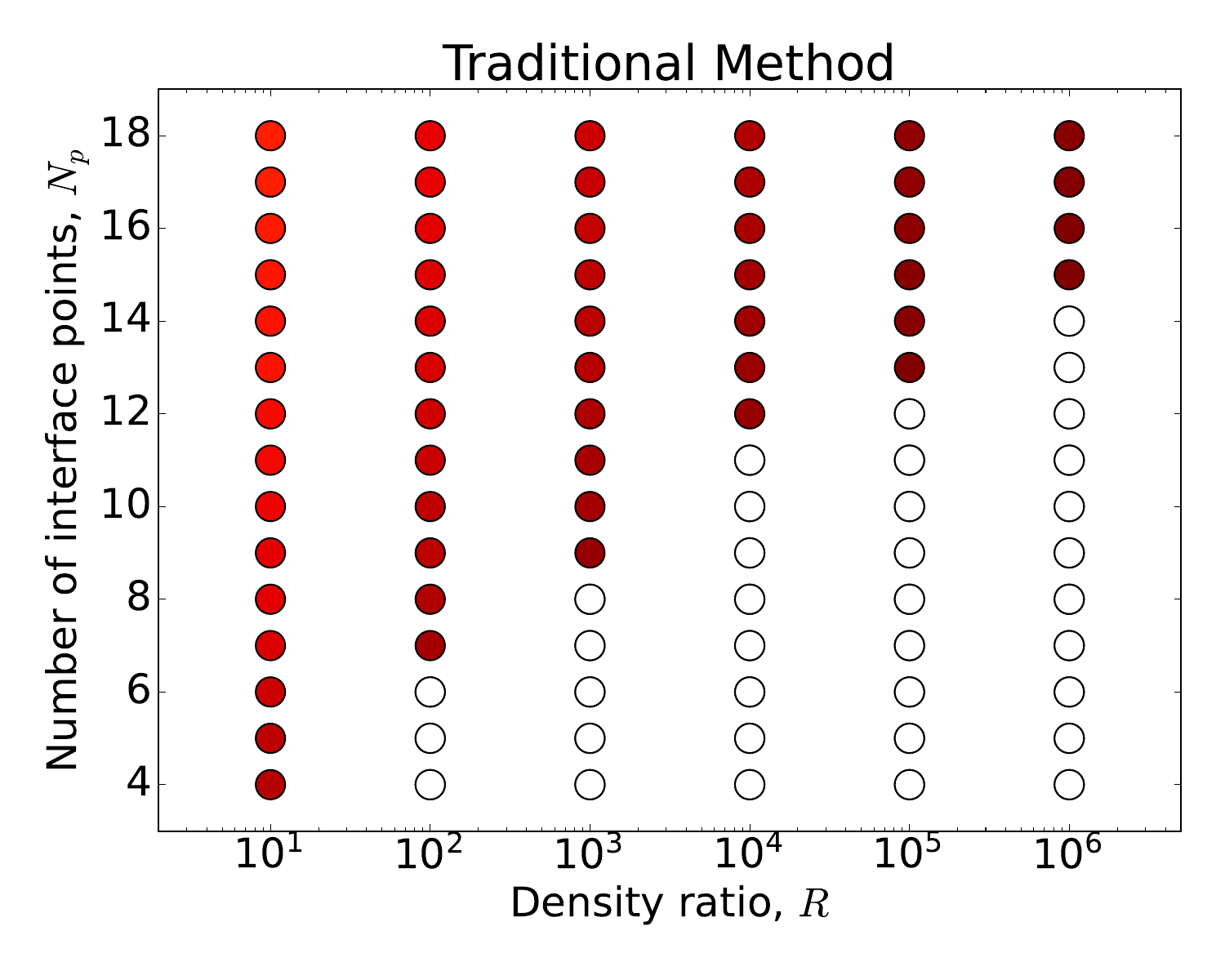}%
    \includegraphics[width=0.3\textwidth,trim={0 0 0 0cm}]{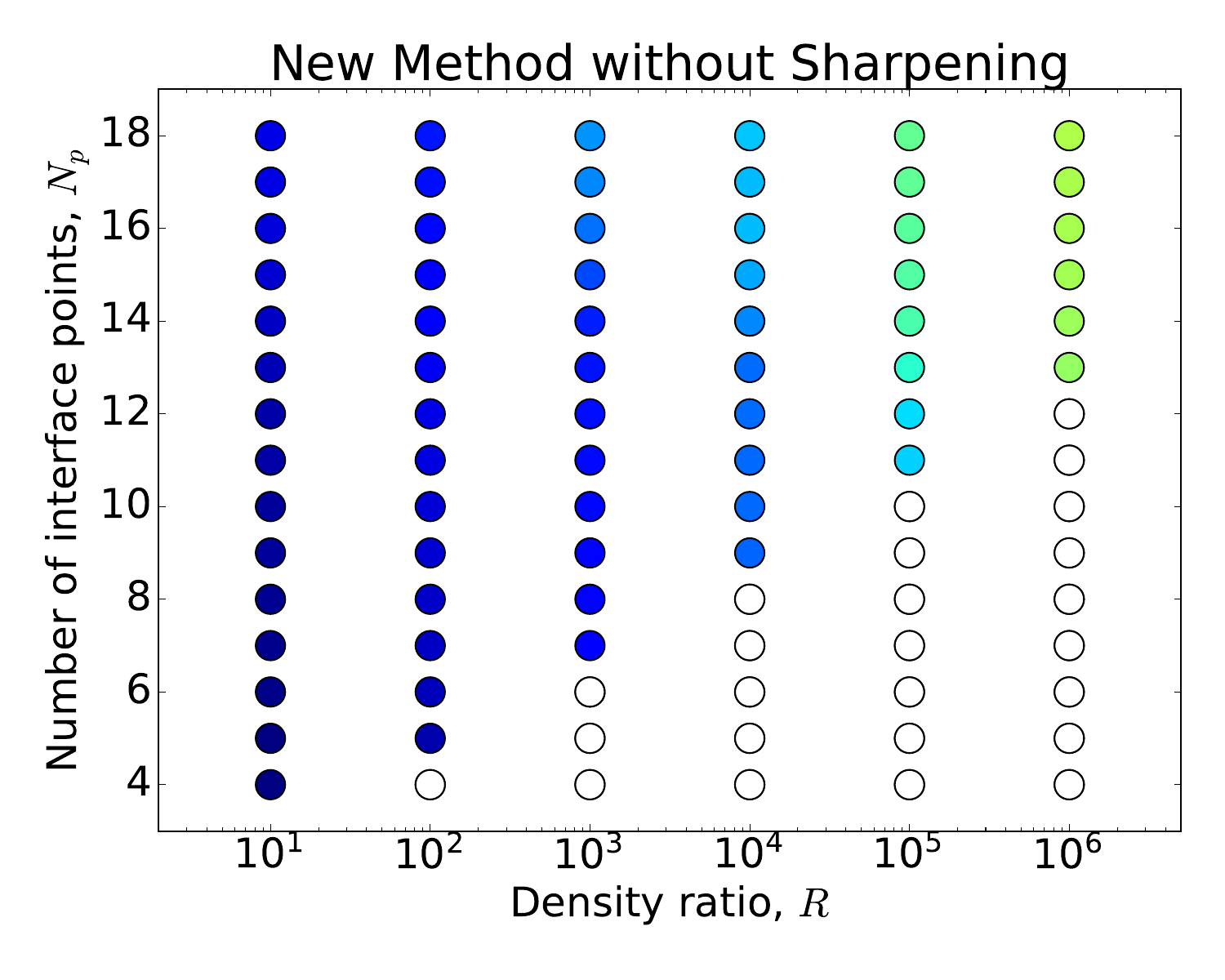}%
    \includegraphics[width=0.3\textwidth,trim={0 0 0 0cm}]{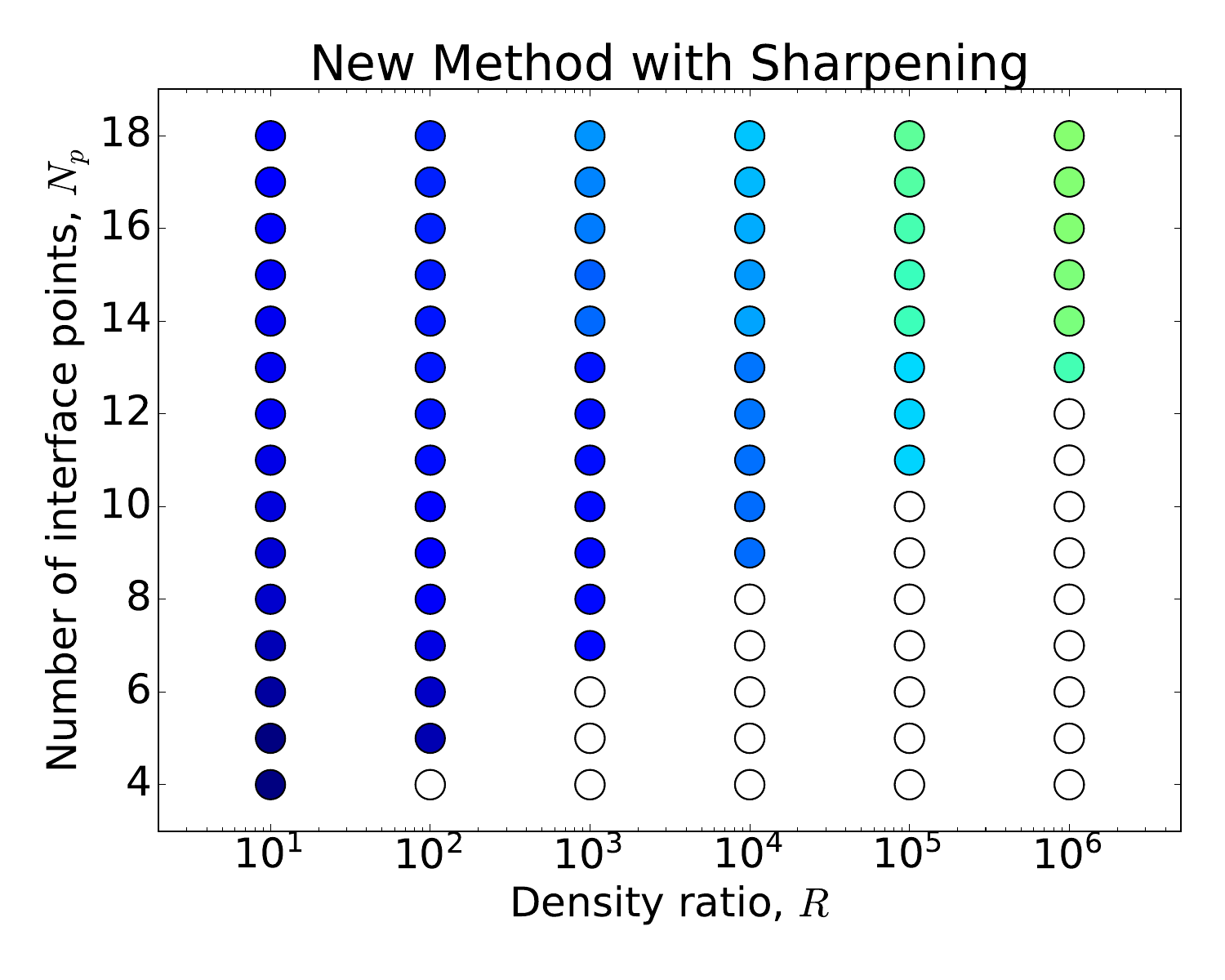}%
    \includegraphics[height=\heightof{\includegraphics[width=0.3\textwidth,trim={0 0 0 0cm}]{figures/stability_compare_no_cbar_new_1-eps-converted-to.pdf}}]{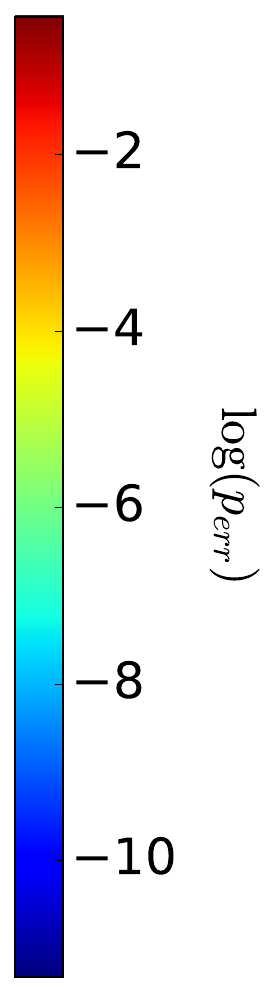}%
    \caption{Pressure oscillation magnitude and stability results for the 2D, three material stability test for a range of number of points in the interface, $N_p$, and maximum density ratios between the materials, $R$. Empty circles indicate that the simulation went unstable before completion. The left plot uses the traditional method, the center plot uses the presented method without sharpening, and the right uses the presented method with sharpening. }
    \label{fig:new_stab}
\end{figure}

First, we repeat the three material stability test introduced in Section~\ref{sec:pokeballtest} with the presented method with and without sharpening. Figure~\ref{fig:new_stab} shows the stability results for the tradition method and the presented method with and without sharpening. For density ratios of $R\geq 100$, stability is seen in the presented method with fewer grid points in the interface, demonstrating the stabilizing effects of the formulation. Figure~\ref{fig:new_stab} also shows that the sharpening term does not affect the stability regimes of the presented LAD method. The pressure error for all cases, with the presented method with and without sharpening is $O(10^{-6}-10^{-12})$ which is 5 to 10 orders of magnitude lower than the traditional method. This stark difference is because the presented scheme maintains equilibrium, while the traditional method does not, as discussed in Sec.~\ref{sec:eqm}.

From these results, it is seen that the presented modifications to the species diffusion flux improve the stability of the compact finite-difference method for large density ratios and reduce the pressure oscillations. Even with the additional sharpening term to maintain sharp interfaces, these benefits are maintained, and the sharpening term improves the stability of the largest density ratios.  

\subsection{Shock-Bubble Interaction}
Now we will examine the performance of the method by studying the classic multi-material test case of a shock interacting with a helium bubble~\cite{jain2023assessment,cook2009enthalpy,huang2023consistent,kawai2011high}. The domain has a nondimensional size $x \in [-2,4]$, $y\in [0,1]$ with wall boundary conditions in $x$ and periodic boundaries in $y$. A bubble of helium is initialized with a radius of $r=25/89$ at $x=0$, $y=0.5$ and air is defined in the rest of the domain. The interface between the two materials is defined with Eq.~\eqref{eqn:srbtanh} with $N_p=14$. The material properties for the air and helium respectively are: $\rho_1 = 1.0$, $\gamma_1 = 1.4$, and $M_1 = 1.0$ and $\rho_2 = 0.138$, $\gamma_2 = 1.67$, and $M_2 = 0.138$. The initial conditions initialize a shock in the air at $x=1$:
\begin{align}
S &= \frac{1}{2}\left(1 + \tanh\left(\frac{x+1}{\frac{3 (14\Delta)}{16}}\right)\right), \\
u &= u_s S, \\ 
v &= 0, \\
\rho &= \rho_s S + \rho_1(1-S), \\
p &= p_s S + p_0 (1-S),
\end{align}
where the states are: $u_s = 0.39473$, $\rho_s = 1.3764$, $p_0 = 1.0$, and $p_s = 1.5698$. For the simulations, the domain is discretized on a uniform Cartesian grid with $N_x=1200$ and $N_y=200$. Sharpening was enabled for the presented method. 

\begin{figure}[]
    \centering
    \begin{subfigure}[b]{0.23\textwidth} 
        \centering
        \includegraphics[width=\textwidth]{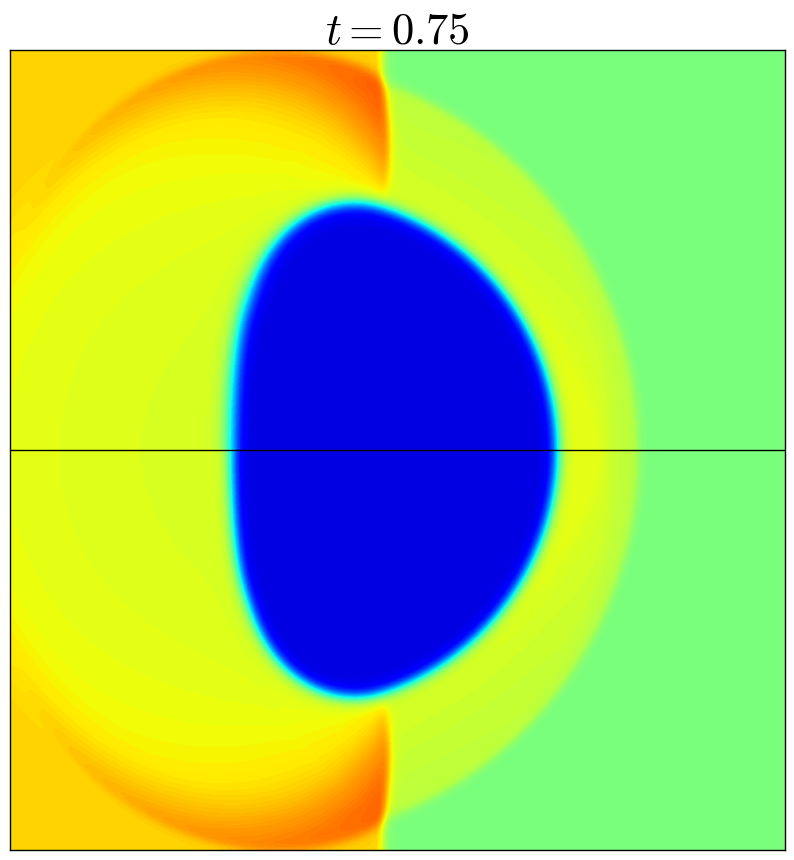}
    \end{subfigure}
    \begin{subfigure}[b]{0.23\textwidth} 
        \centering
        \includegraphics[width=\textwidth]{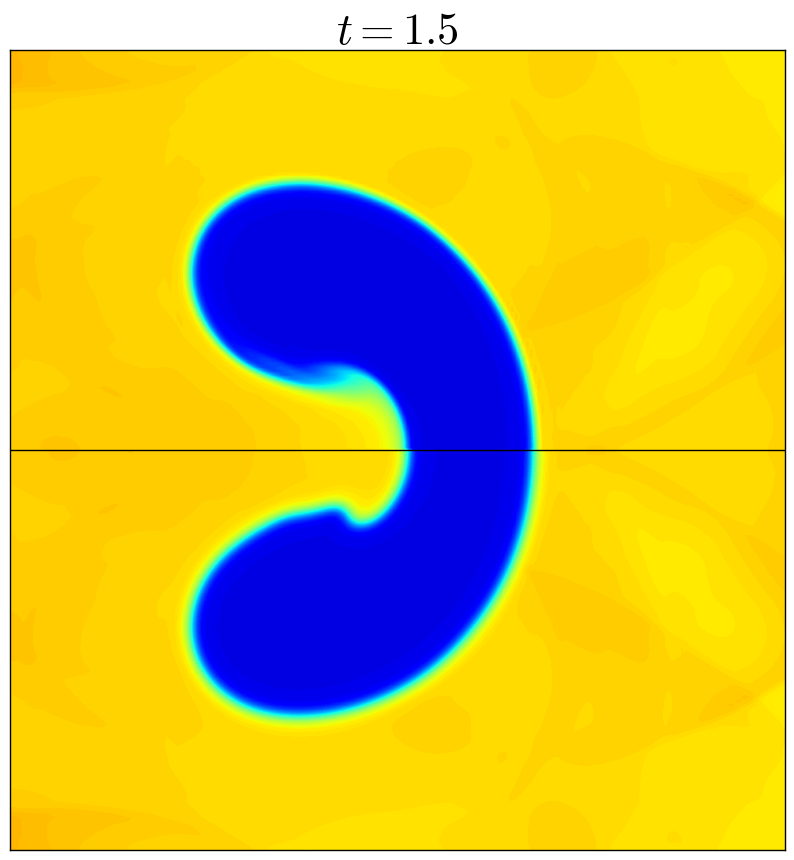}
    \end{subfigure}
    \begin{subfigure}[b]{0.23\textwidth} 
        \centering
        \includegraphics[width=\textwidth]{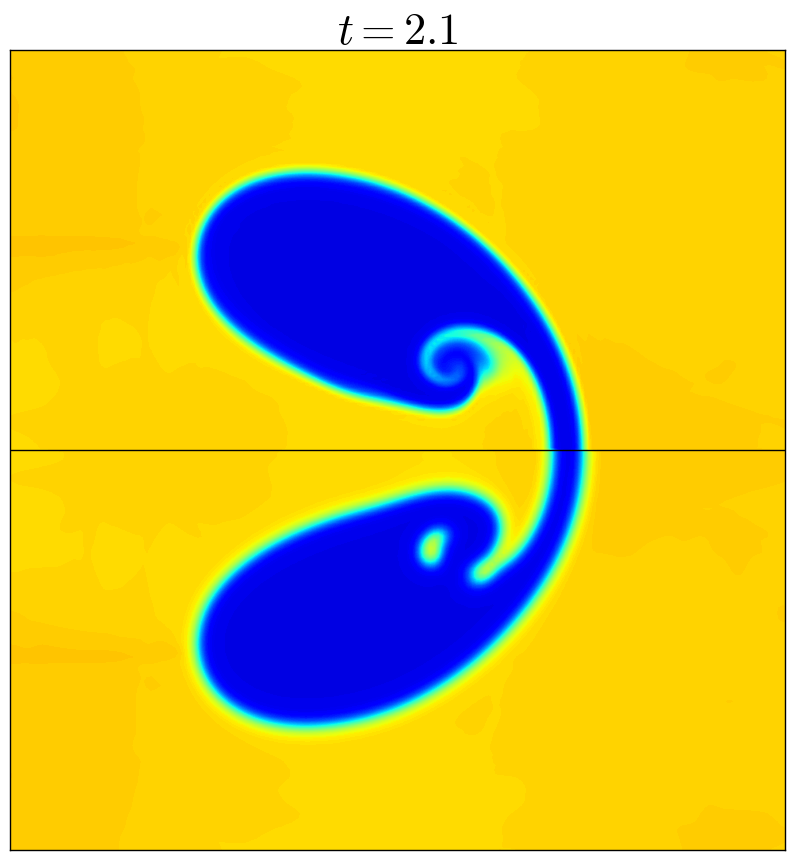}
    \end{subfigure}
    \begin{subfigure}[b]{0.23\textwidth} 
        \centering
        \includegraphics[width=\textwidth]{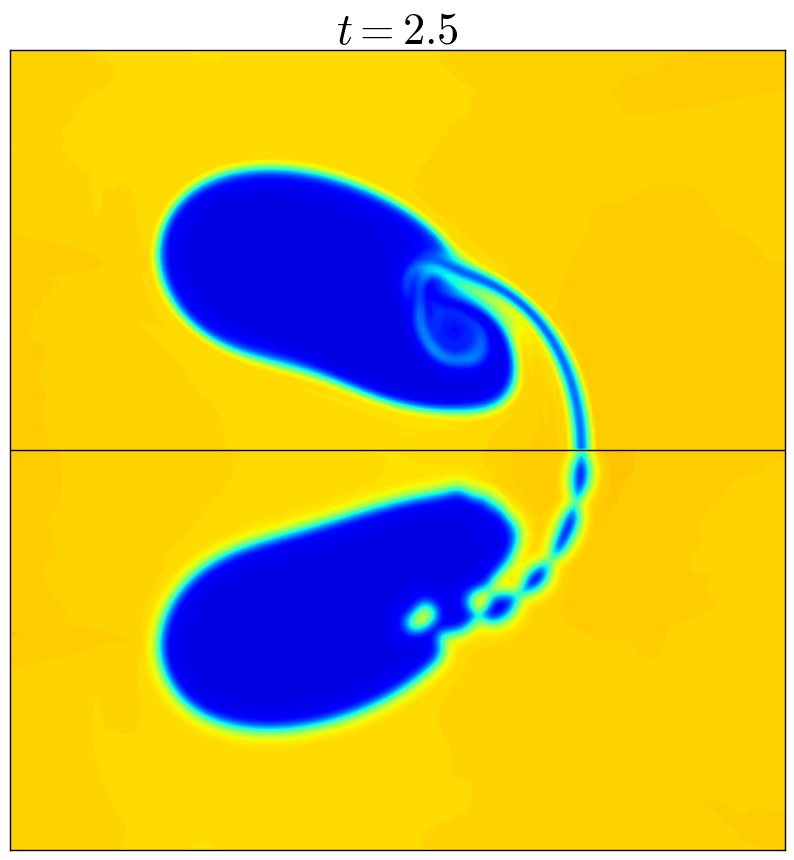}
    \end{subfigure}

    \begin{subfigure}[b]{0.23\textwidth} 
        \centering
        \includegraphics[width=\textwidth]{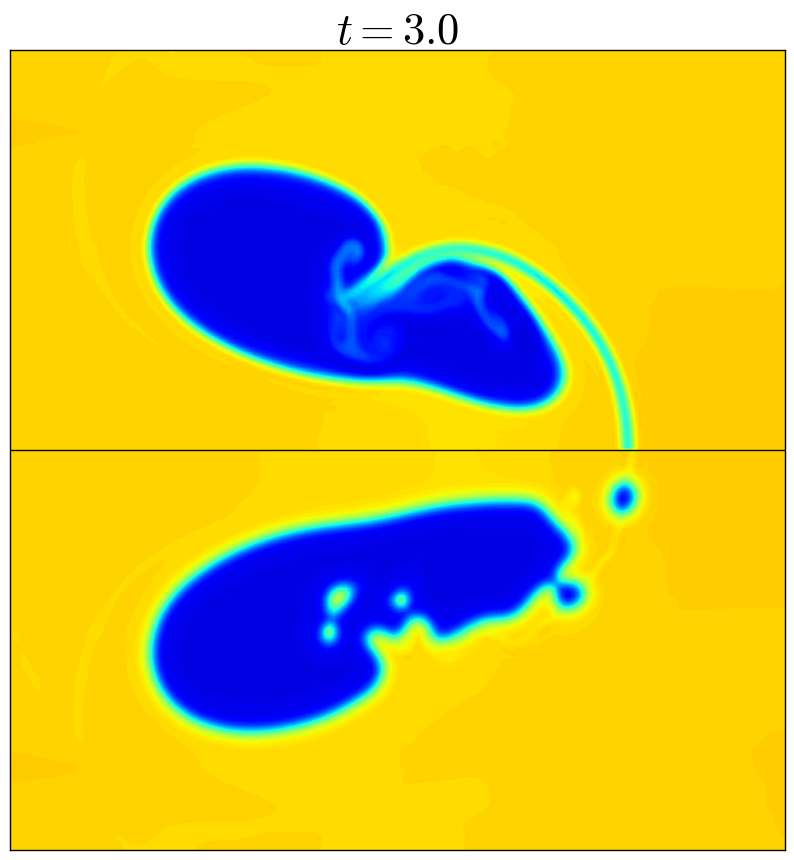}
    \end{subfigure}
    \begin{subfigure}[b]{0.23\textwidth} 
        \centering
        \includegraphics[width=\textwidth]{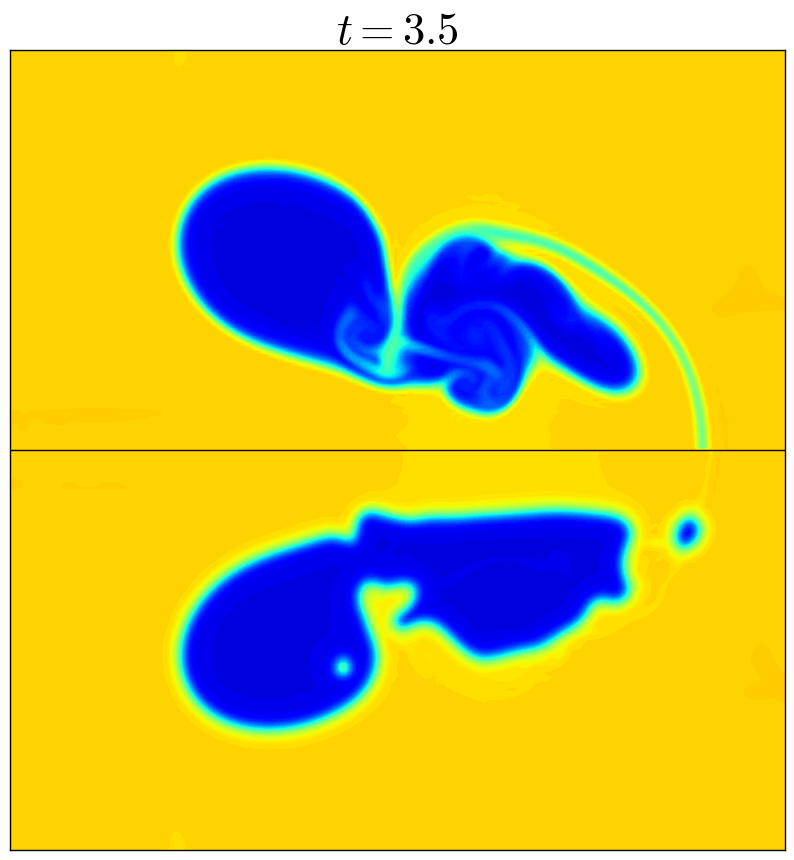}
    \end{subfigure}
    \begin{subfigure}[b]{0.23\textwidth} 
        \centering
        \includegraphics[width=\textwidth]{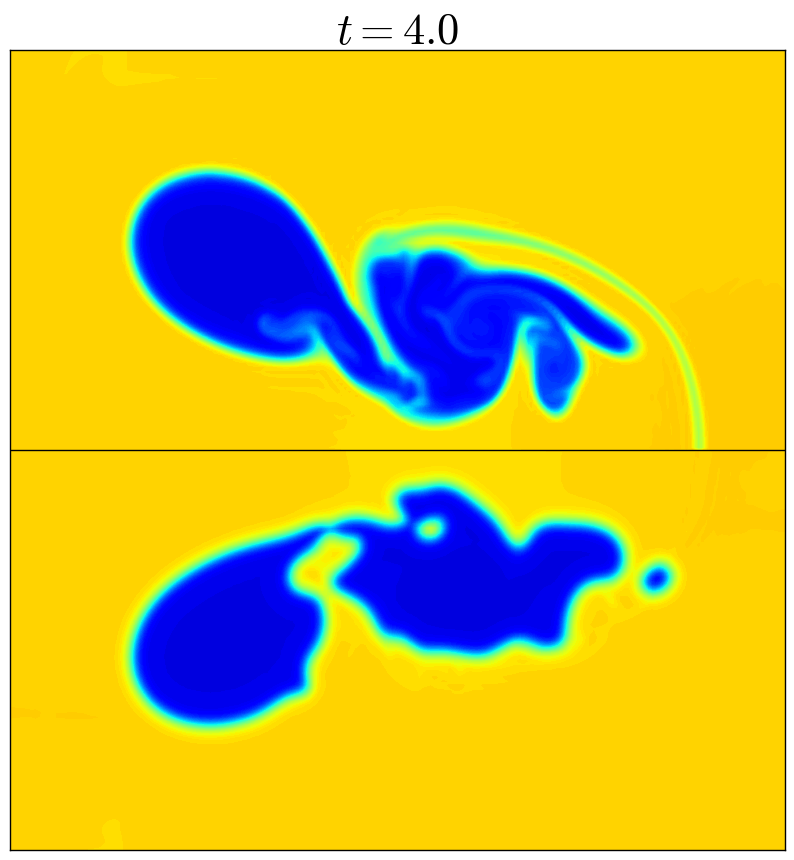}
    \end{subfigure}
    \begin{subfigure}[b]{0.23\textwidth} 
        \centering
        \includegraphics[width=\textwidth]{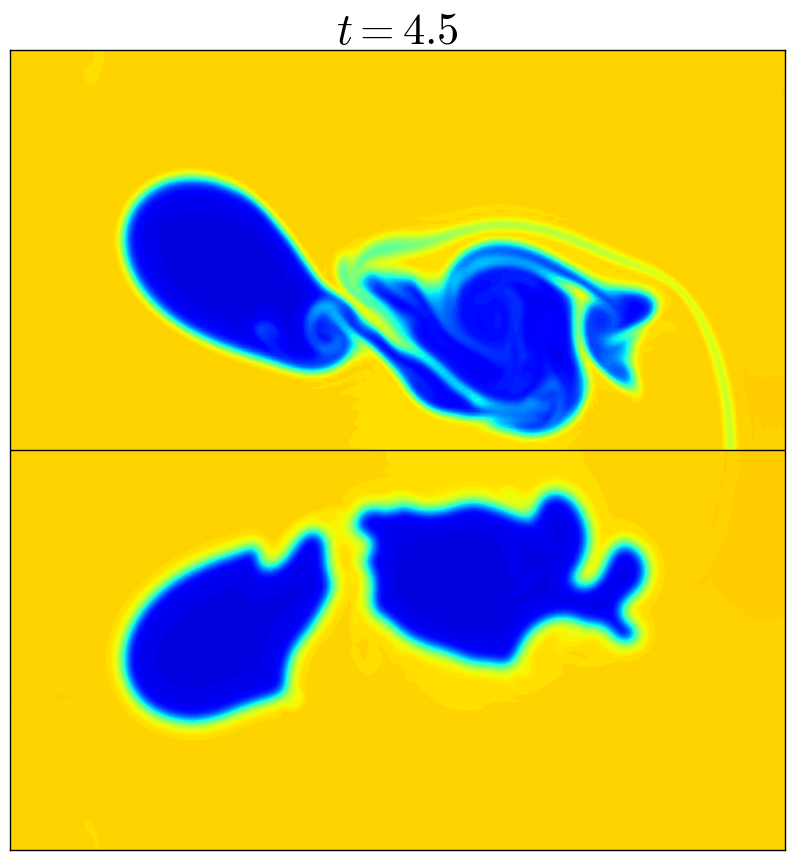}
    \end{subfigure}

    \begin{subfigure}[b]{0.75\textwidth} 
        \centering
        \includegraphics[width=\textwidth]{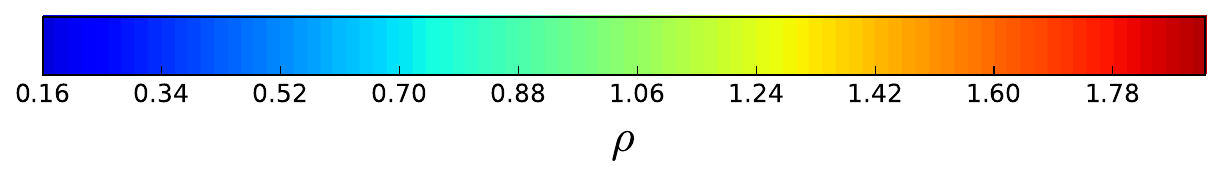}
    \end{subfigure}
    \caption{Comparisons between the traditional artificial diffusivity method (top) and the presented method with sharpening enabled (bottom) for a shock in air causing the breakup of a helium bubble.}
    \label{fig:he_bubble}
\end{figure}

For this test case, the density ratio is only $7.25$, so the improved stability from the presented method is not necessary. Additionally, for the case of helium and air, the gases diffuse into one another, so a sharp interface may not be the best way to model these materials. Instead, this test case serves to demonstrate the effect of the interface sharpening on a well-studied test case. Figure~\ref{fig:he_bubble} shows the results at different snapshots in time for both the traditional method and the presented method. Using the presented method, sharp interfaces of uniform thickness are seen throughout the simulation. As the jet reaches the far side of the bubble, a thin filament forms. For the traditional method, this filament persists throughout the simulation as a region of low volume fraction helium. For the presented method, as the filament gets thin, it breaks into small helium bubbles, which later rejoin the larger bubbles on the top and bottom. The sharpening term causes these bubbles to form, because it sharpens the interfaces in mixed regions. The sharpening term sharpens to a thickness based on the $\epsilon$ parameter from Eq.~\eqref{eqn:sharp_term} which is a function of the grid size. Hence, the size of these bubbles is based on the numerical grid size, not any physical properties. Experimentally, it was observed in this simulation and others that the sharpening term creates bubbles of radius of approximately $5\Delta$. Hence, the size of the bubbles can be reduced with mesh refinement. In this case, the sharp interface of the presented method is not expected to be more representative of the true physics, but this case serves as a good demonstration of the properties that the method displays.

\begin{figure}[]
    \centering
    \begin{subfigure}[b]{0.23\textwidth} 
        \centering
        \includegraphics[width=\textwidth]{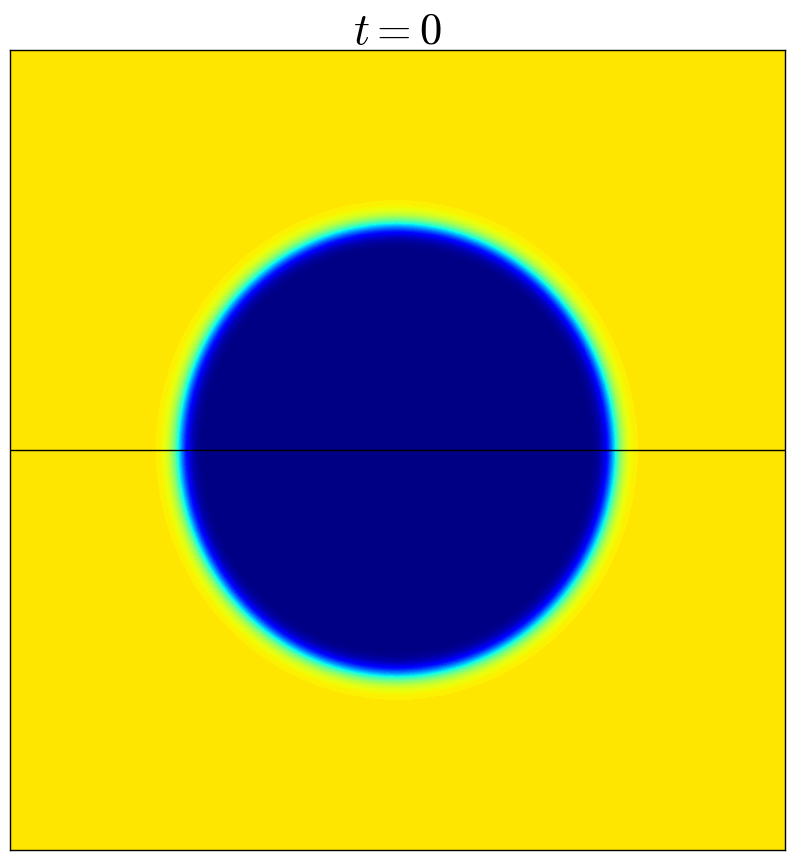}
    \end{subfigure}
    \begin{subfigure}[b]{0.23\textwidth} 
        \centering
        \includegraphics[width=\textwidth]{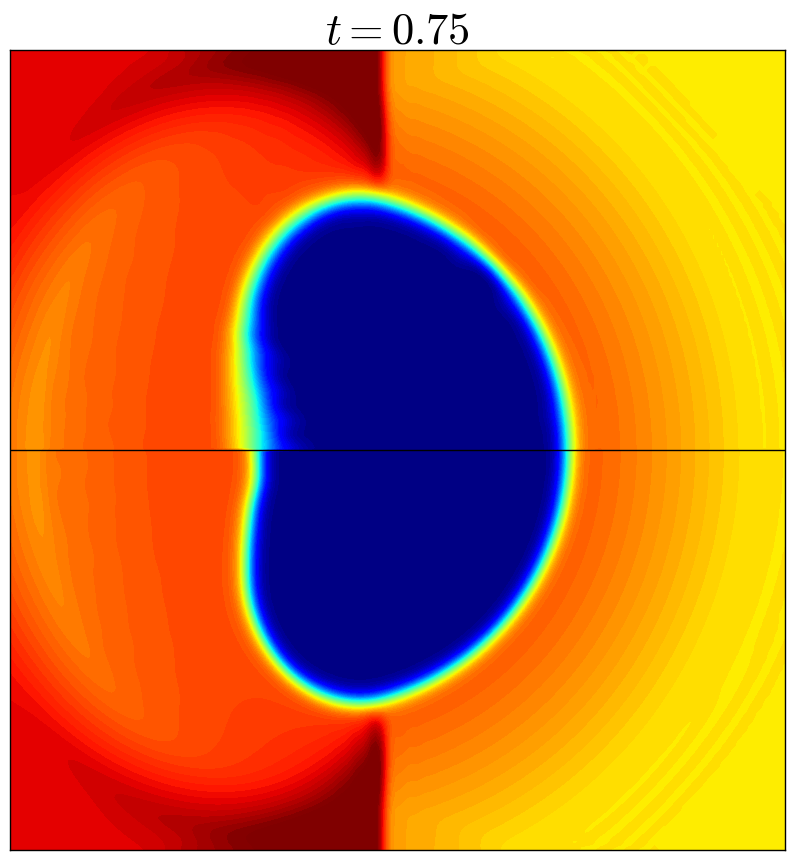}
    \end{subfigure}
    \begin{subfigure}[b]{0.23\textwidth} 
        \centering
        \includegraphics[width=\textwidth]{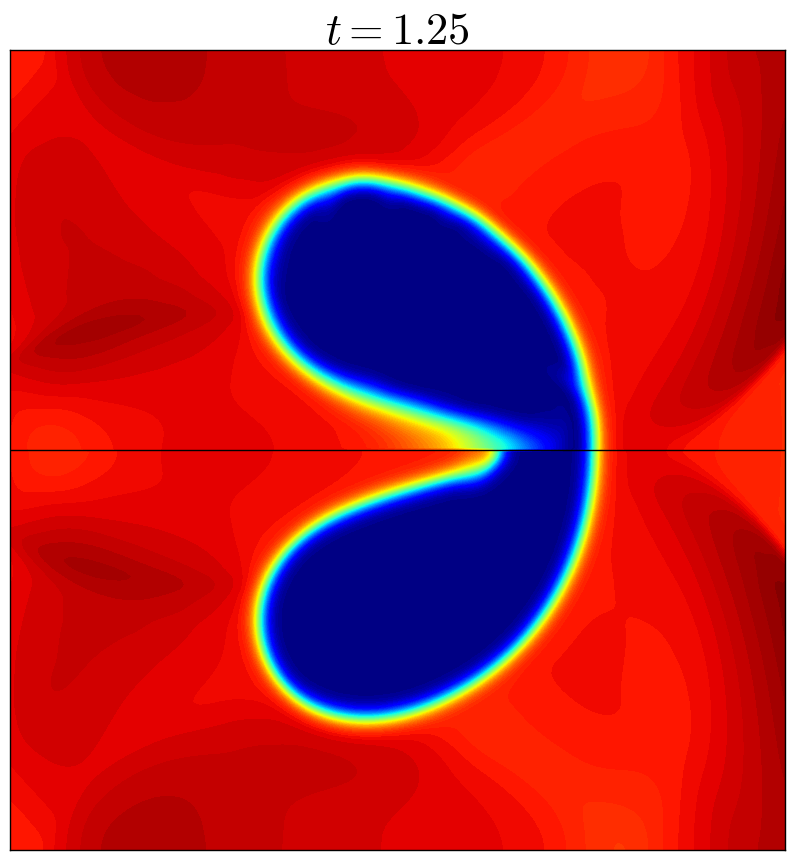}
    \end{subfigure}
    \begin{subfigure}[b]{0.23\textwidth} 
        \centering
        \includegraphics[width=\textwidth]{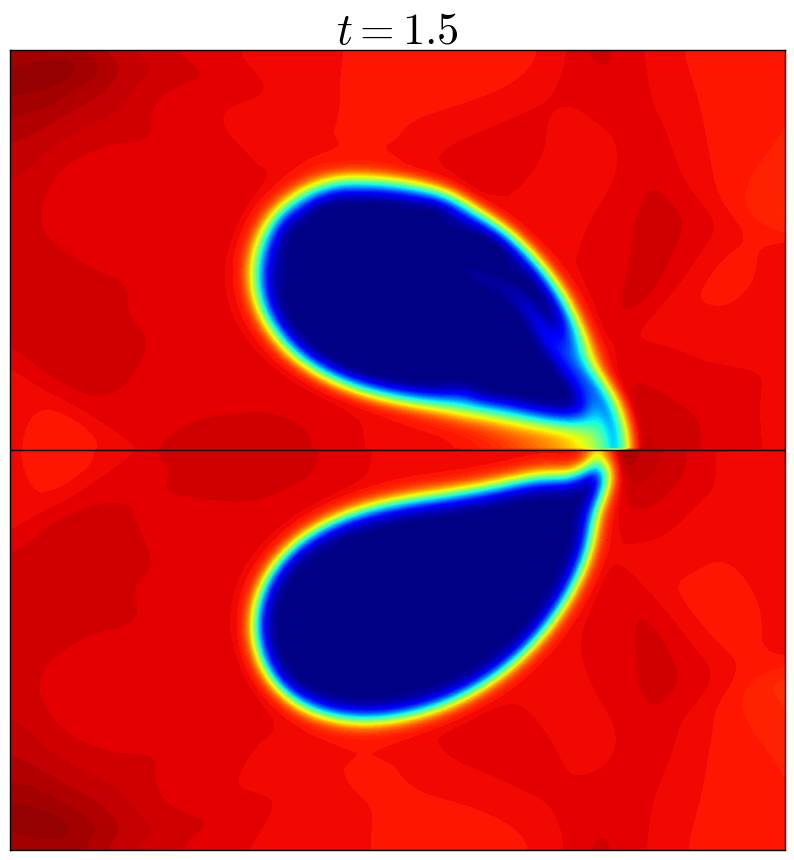}
    \end{subfigure}

    \begin{subfigure}[b]{0.23\textwidth} 
        \centering
        \includegraphics[width=\textwidth]{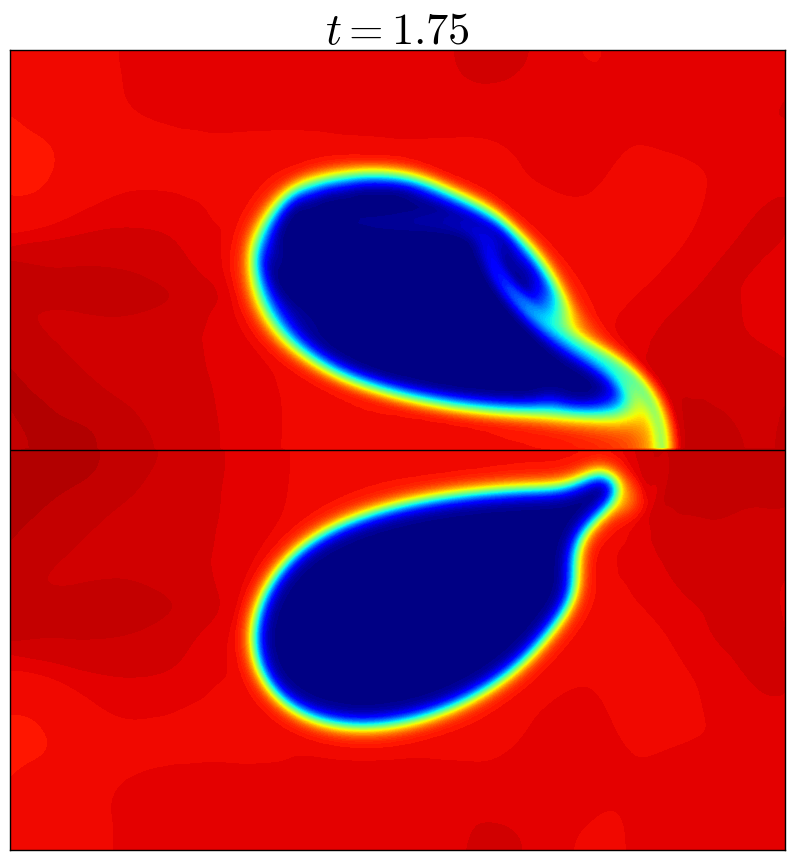}
    \end{subfigure}
    \begin{subfigure}[b]{0.23\textwidth} 
        \centering
        \includegraphics[width=\textwidth]{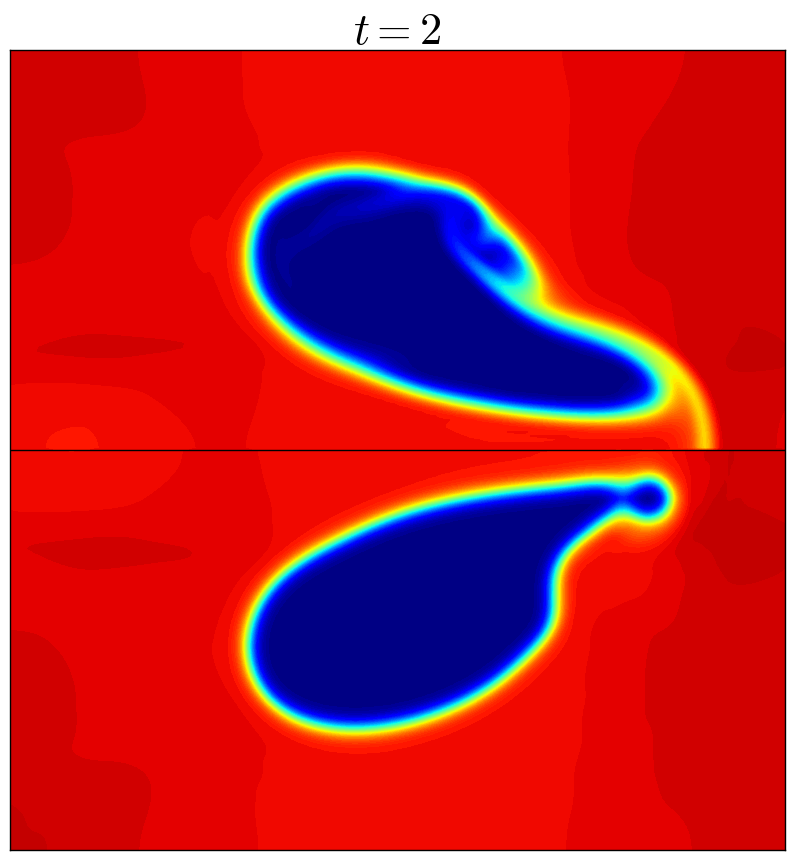}
    \end{subfigure}
    \begin{subfigure}[b]{0.23\textwidth} 
        \centering
        \includegraphics[width=\textwidth]{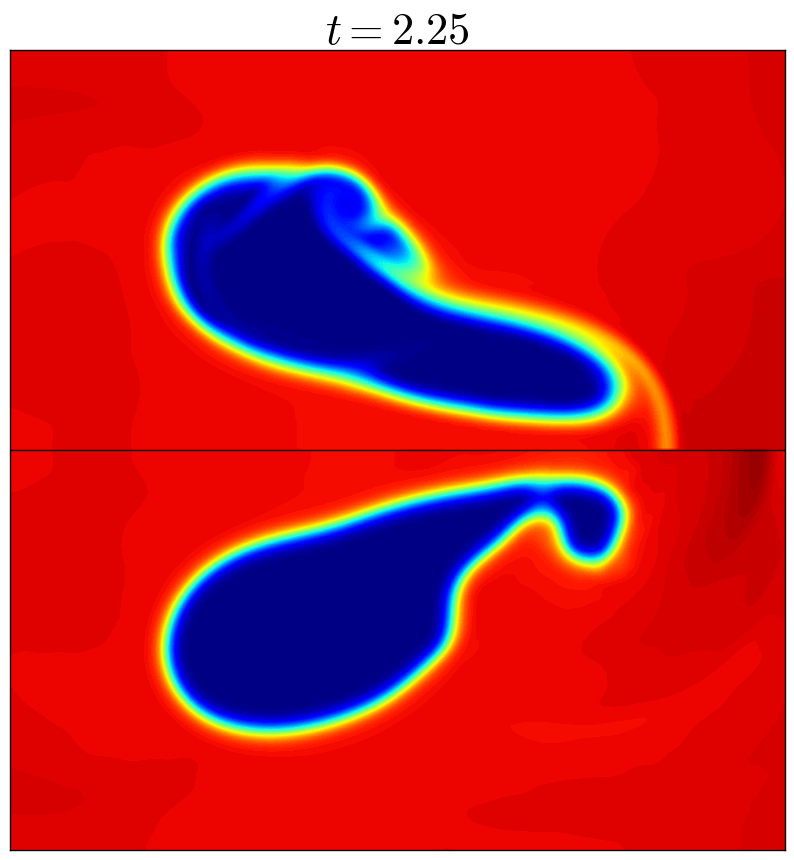}
    \end{subfigure}
    \begin{subfigure}[b]{0.23\textwidth} 
        \centering
        \includegraphics[width=\textwidth]{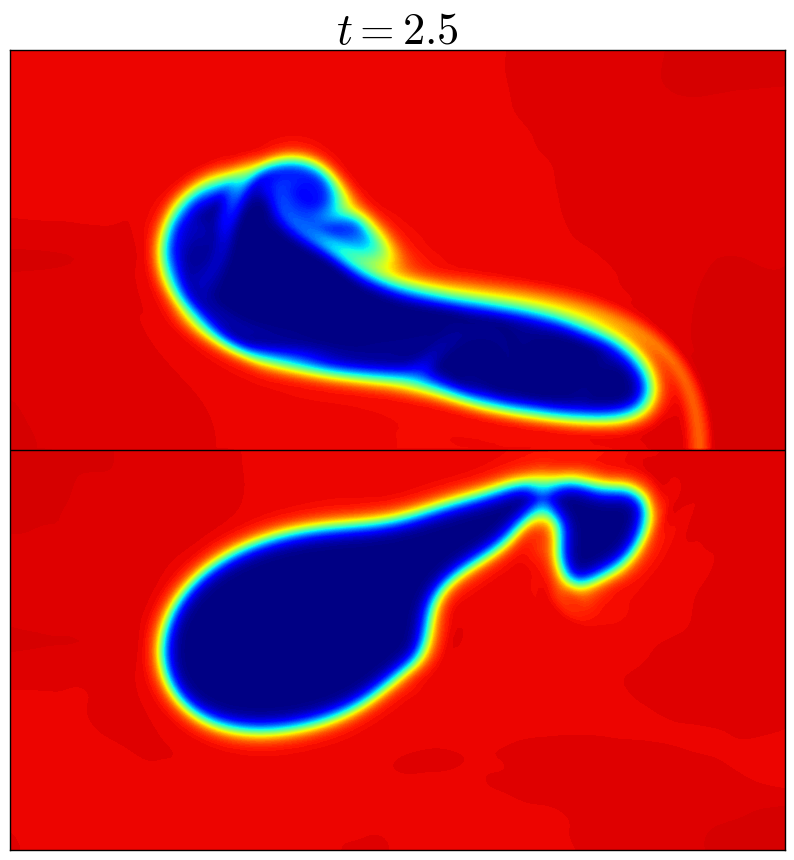}
    \end{subfigure}

    \begin{subfigure}[b]{0.75\textwidth} 
        \centering
        \includegraphics[width=\textwidth]{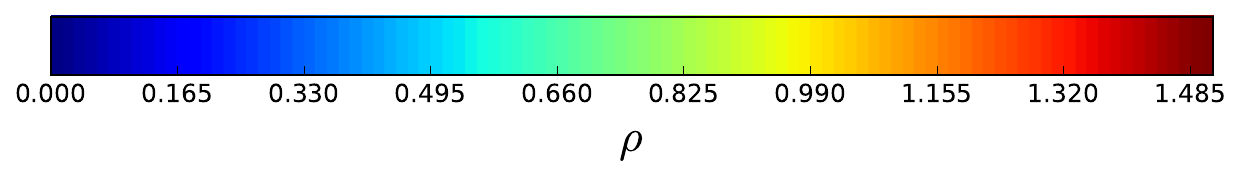}
    \end{subfigure}
    \caption{Comparisons between the presented method without sharpening (top) and with sharpening enabled (bottom) for a shock in air causing the breakup of a helium bubble with density $\rho=0.138\times10^{-5}$. This case is unstable for the traditional LAD method.}
    \label{fig:high_rho_he_bubble}
\end{figure}

In order to highlight the benefits of the presented method for large density ratio flows, we repeated the shock-bubble interaction case, but decreased the density of helium by $5$ orders of magnitude, so the material properties are $\rho_1 = 1.0$, $\gamma_1 = 1.4$, and $M_1 = 1.0$ and $\rho_2 = 1.38\times 10^{-6}$, $\gamma_2 = 1.67$, and $M_2 = 1.38\times 10^{-6}$. For this case, the density ratio is about $725000$. The artificial diffusivity coefficients are $C_\mu=10^{-3}$, $C_\beta = 0.7$, $C_{D,Y}=C_{D,V}=10^{-2}$, and $\epsilon = 1.2\Delta$. This larger value of $\epsilon$ was observed to better stabilized this large density ratio in the prescence of the shock and had minimal effect on the thickness of the interface. As expected from the stability test, the traditional method was unstable for this large of a density ratio while the presented method is able to run stably. Increasing the diffusivity coefficients and decreasing the CFL number were unable to stabilize the traditional method. Figure~\ref{fig:high_rho_he_bubble} shows the results for the presented method with and without sharpening at various times. With such a low density in the bubble, the shock quickly passes through it and jetting causes the bubble to break into two stretched halves. Once again, when the sharpening term is used, small bubbles form as the jet is breaking through the far side of the bubble. While this case does not have any physical relevance, it shows that the presented method is able to remain stable for large density ratios with and without sharpening in cases with shocks.

\subsection{Rayleigh-Taylor Instability}
\begin{figure}[]
    \centering
    \begin{subfigure}[b]{0.45\textwidth} 
        \centering
        \includegraphics[width=\textwidth]{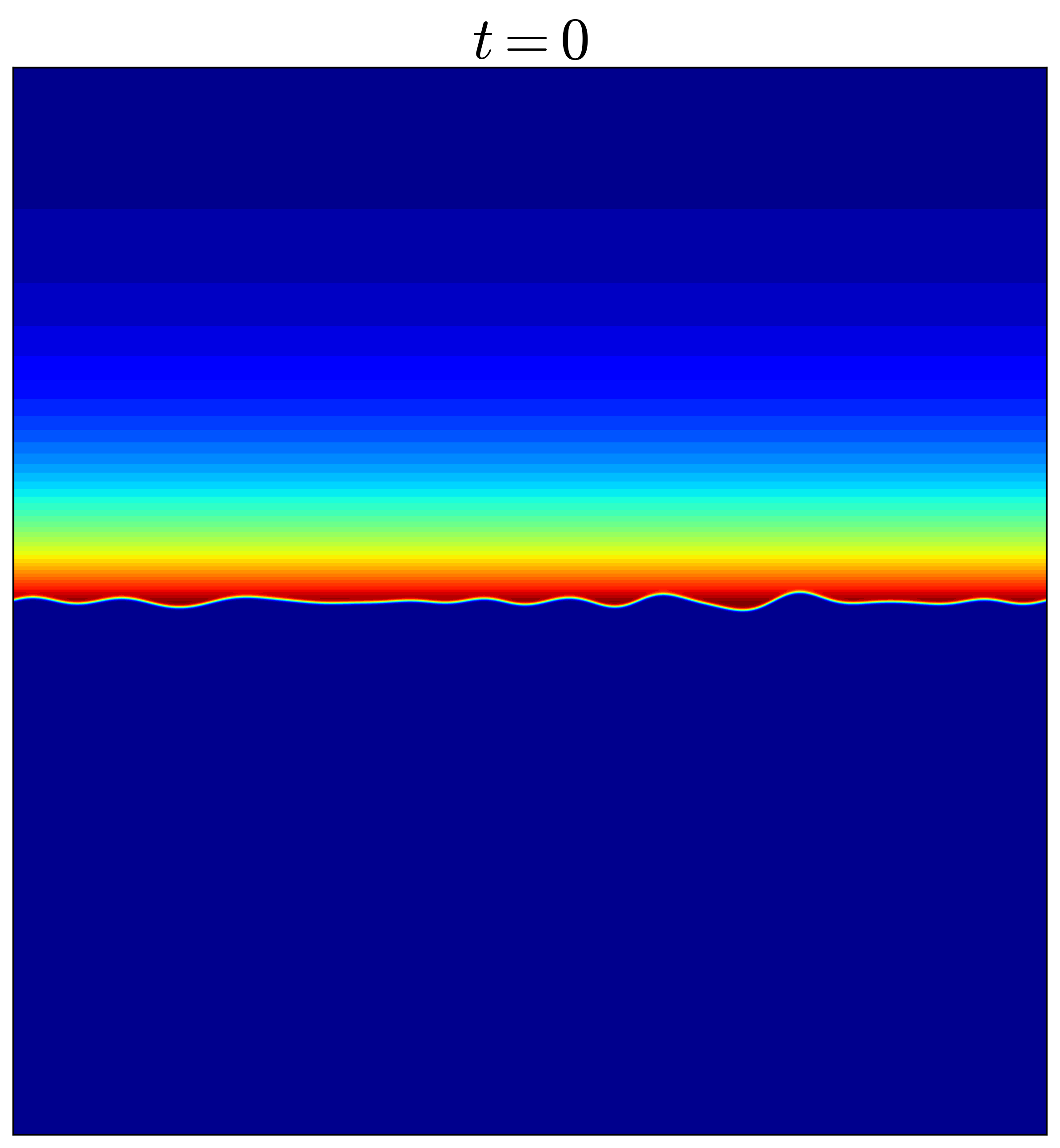}
        \centering
    \end{subfigure}
    \begin{subfigure}[b]{0.45\textwidth} 
        \centering
        \includegraphics[width=\textwidth]{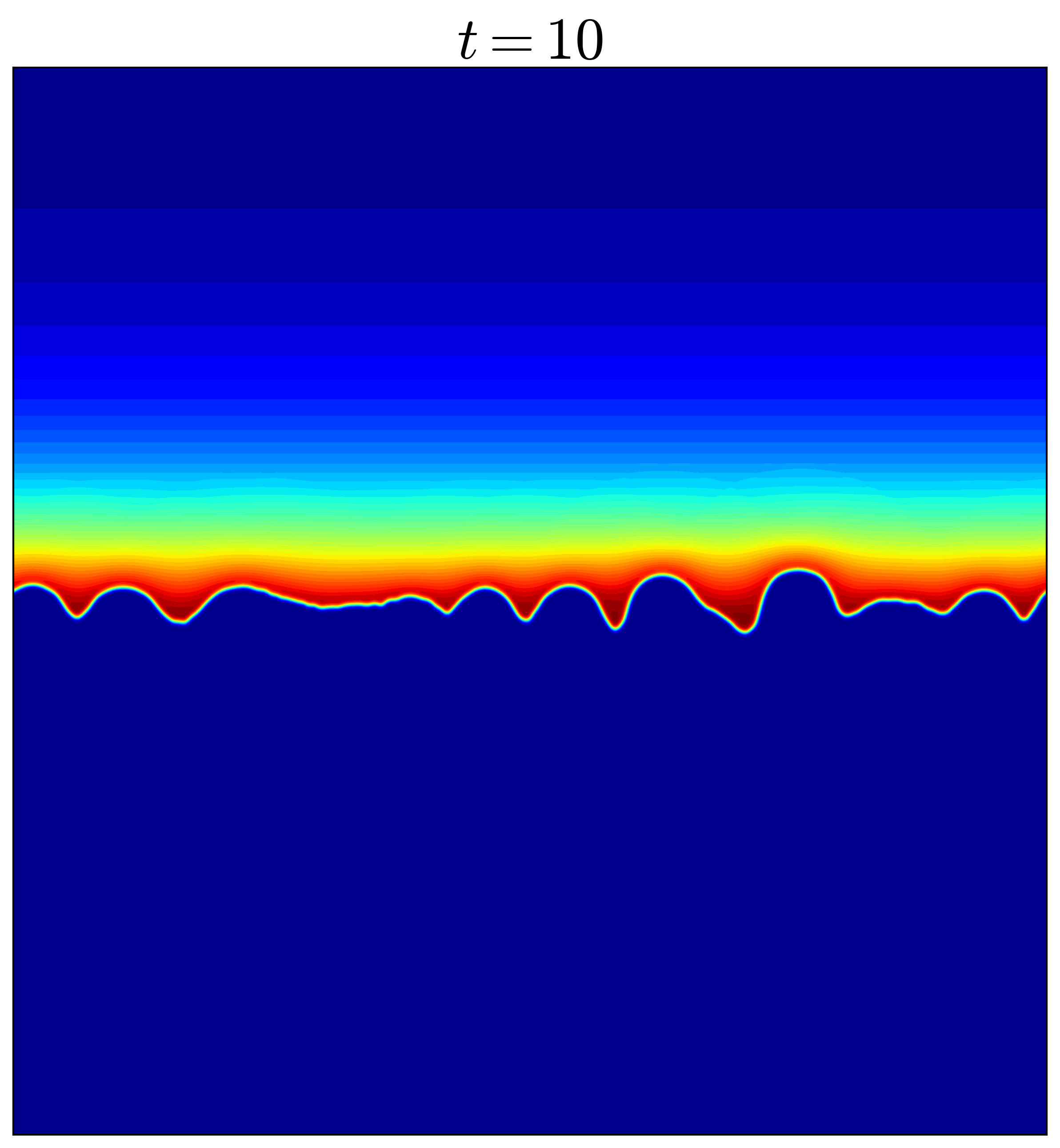}
    \end{subfigure}

    \begin{subfigure}[b]{0.45\textwidth} 
        \centering
        \includegraphics[width=\textwidth]{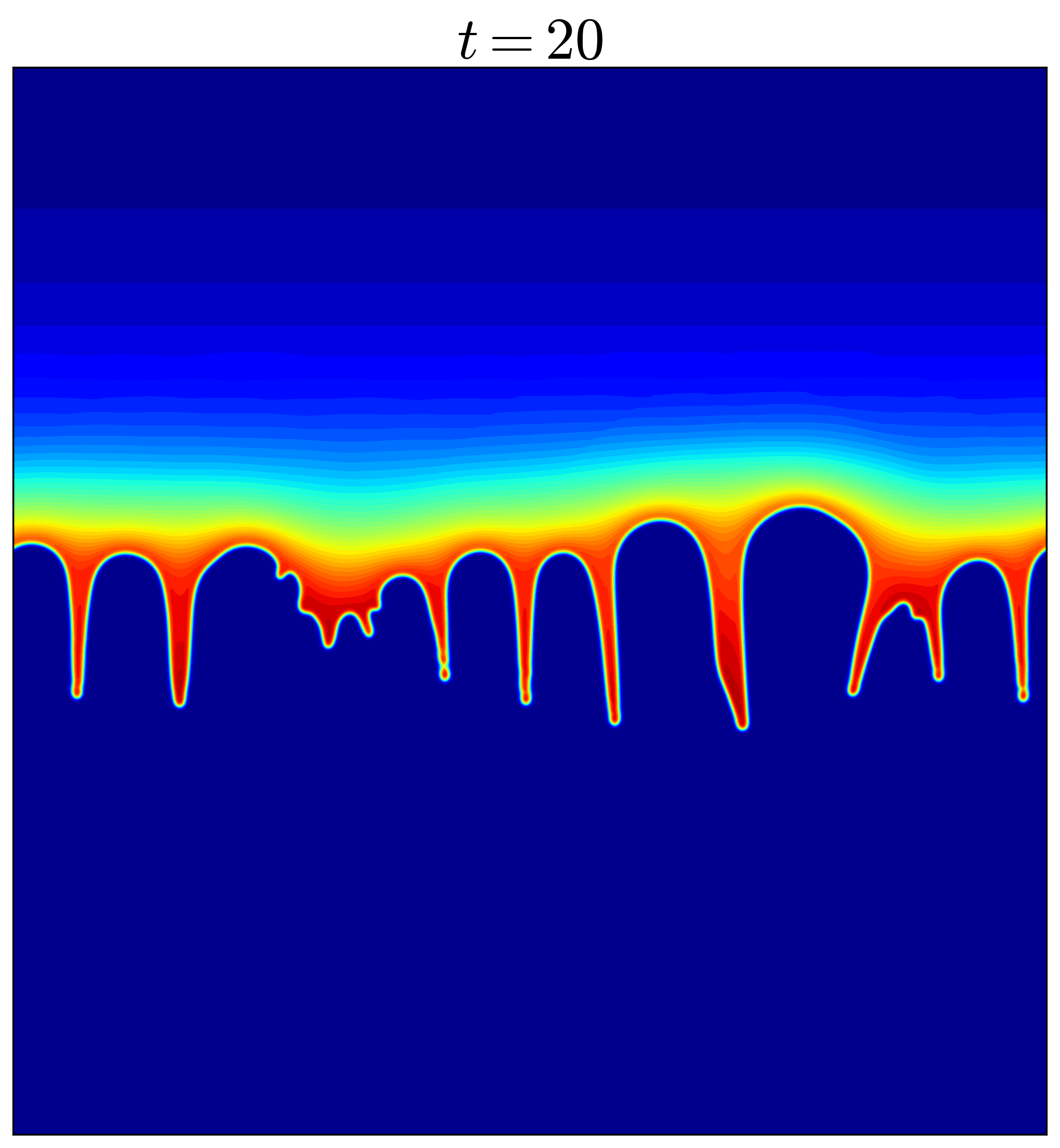}
    \end{subfigure}
    \begin{subfigure}[b]{0.45\textwidth} 
        \centering
        \includegraphics[width=\textwidth]{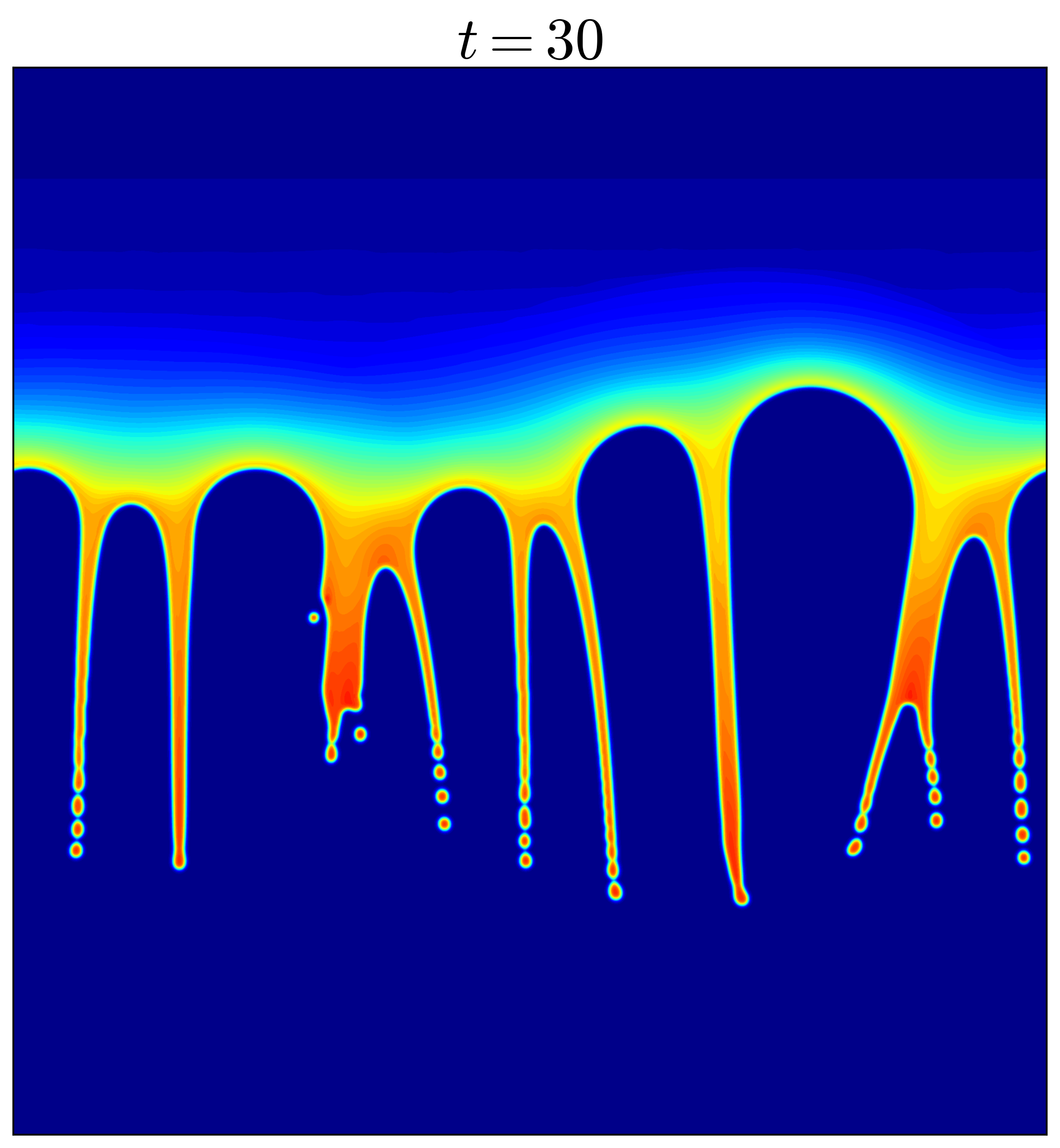}
    \end{subfigure}

    \begin{subfigure}[b]{0.75\textwidth} 
        \centering
        \includegraphics[width=\textwidth]{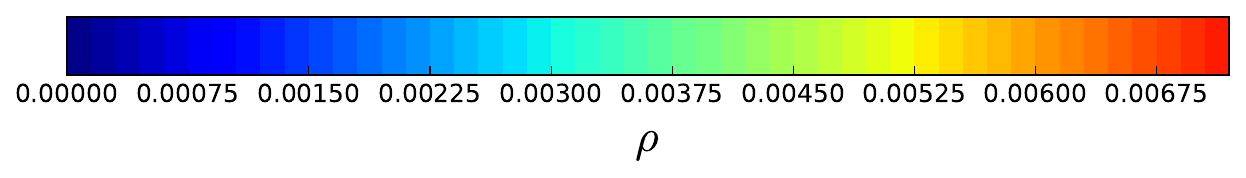}
    \end{subfigure}
    \caption{Plots of density of $At=0.9996$ Rayleigh-Taylor instability stabilized with the presented LAD method. This simulation is unstable when run with the traditional LAD method.}
    \label{fig:rt}
\end{figure}

In order to demonstrate the high-density ratio benefits on a more physically relevant problem, we simulated a high-density ratio Rayleigh-Taylor instability. A $2D$ domain of size $[-64,64]\times[-64,64]\,\text{cm}$ with periodic boundaries in $x$ and wall boundaries in $y$ is discretized with $N=1000$ grid points in each direction. A gravitational acceleration of $-980\,\text{cm}/\text{s}^2$ is applied in the $y$ direction. A heavy fluid is initialized above a light fluid with a randomly perturbed interface centered around $y=0\,\text{cm}$ defined as
\begin{align}
    V_h(x,y,t=0) &= \frac{1}{2}\left(1-\text{erf}\left(\frac{y-\xi(x)}{2\Delta}\right)\right), 
\end{align}
where $\xi(x)$ are the random perturbations following a Gaussian distribution in wavenumber space with a modal peak of the perturbation spectrum of $k_0=8$, modal standard deviation of the perturbation spectrum of $\sigma=4$, and root-mean-square of the perturbations of 
\begin{align}
\left[2\int_{0}^{k_{\max}} E_\xi(k)\,\text{d}k\right]^{1/2} = E_{rms},
\end{align} 
where $k_{\max} = N/3-1$ and $E_{rms} = 4000\,\text{cm}$. The materials are $\gamma$-law gases with molecular masses of $M_l= 4.002602\times 10^{-2}\,\text{g}/\text{mol}$ and $M_h=200.1301\,\text{g}/\text{mol}$ and $\gamma_l=\gamma_h=1.4$ for the light and heavy materials respectively. The resulting density ratio is $5000$ and the Atwood number is $0.9996$. The density in the domain is stratified so the materials are at hydrostatic equilibrium with interface pressure $p=10^6\,\text{Ba}$ at $T=300\,\text{K}$. The artificial diffusivity parameters are $C_\mu=4e-3$, $C_\beta = 1e-1$, and $C_{D,Y}=C_{D,V}=2e-2$. Figure~\ref{fig:rt} shows the density field of the flow at various times throughout the simulation. As expected with Rayleigh-Taylor simulations, bubbles of the light material move upwards and spikes of the heavy material move downward with gravity. As the spikes accelerate and get thin, the sharpening term causes small droplets to separate, as was seen in the other cases. The traditional method was unable to stabilize simulations with this large Atwood number, showing the strength of the presented method. This case also demonstrates how the sharpening term enables the study of immiscible Rayleigh-Taylor flows, but a detailed analysis of those cases is outside of the scope of this work~\cite{olson2024large}.

\subsection{Droplets in Shear Flow} \label{sec:shearflow}
Following the work of Mirjalili et.\ al.~\cite{mirjalili2024conservative}, we extend the standard drop in a shear flow test~\cite{rider1995stretching} to a three-material flow with two drops of different materials. This case will test that the method can maintain sharp interfaces under shearing and that it properly handles $N$-material interfaces. The domain has size $x \in [0,1]$ and $y\in [0,1]$ with periodic boundaries. Two circular droplets with radius $r=0.15$ of materials $1$ and $2$ are initialized at $(0.75,0.5)$ and $(0.5,0.75)$ respectively. The background material is material $3$. All three materials has the same properties: $\gamma = 1.4$, $\rho = 1.293\times 10^{-3}$, and $M=1$. The velocity field is prescribed by the stream function:
\begin{align}
    \Psi(x,y,t) = \frac{1}{\pi}\sin^2{(\pi x)}\sin^2{(\pi y)}\cos{\left(\frac{\pi t}{T}\right)},
\end{align}
where $T=4$ is the period of the vortex. The prescribed velocity will shear the droplets and then the vortex will reverse and return the droplets to their original positions. Analytically, we expect the solution at $t=T$ to be the same as $t=0$. 

\begin{figure}[]
    \centering
    \begin{subfigure}[b]{0.45\textwidth} 
        \centering
        \includegraphics[width=\textwidth]{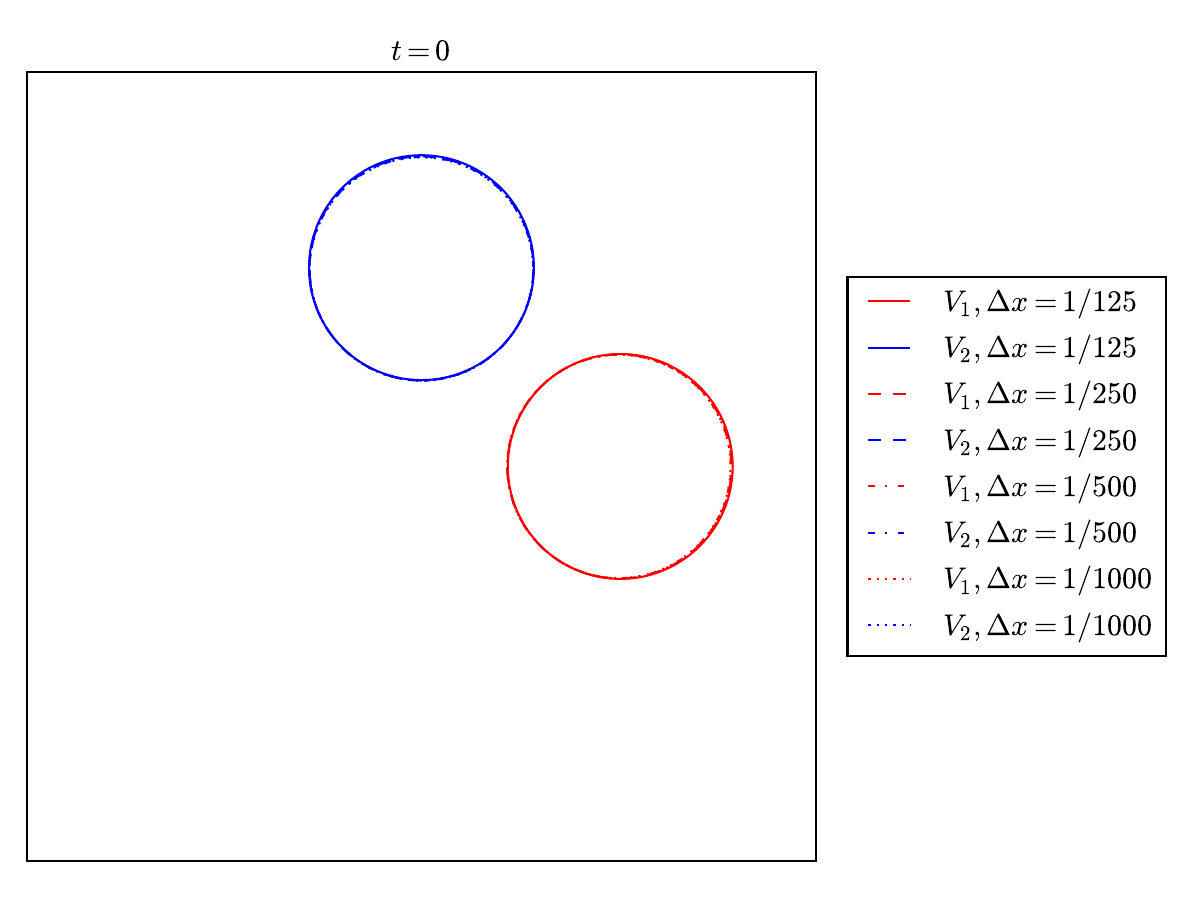}
        \centering
        \caption{ }
        \label{fig:vortex_ic}
    \end{subfigure}
    \begin{subfigure}[b]{0.45\textwidth} 
        \centering
        \includegraphics[width=\textwidth]{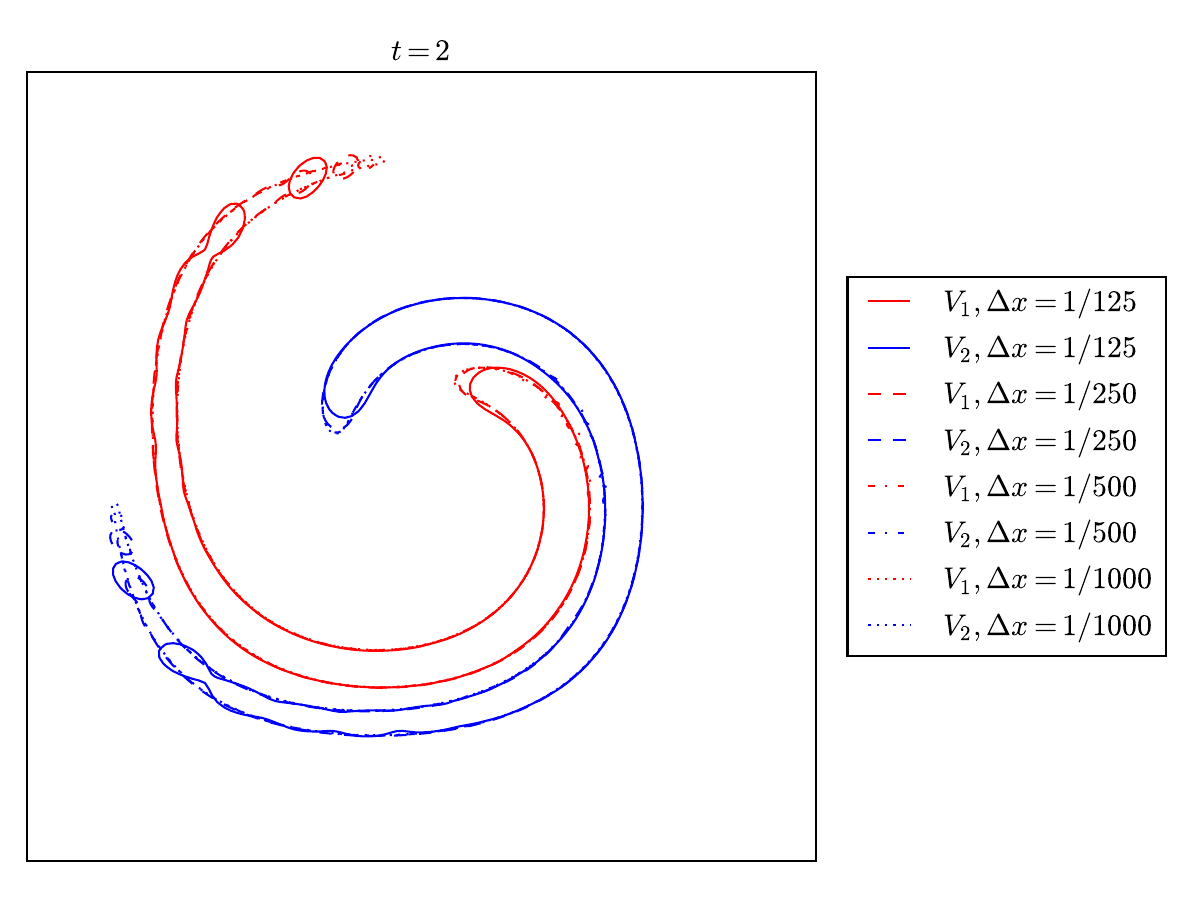}
        \caption{ }
        \label{fig:vortex_mid}
    \end{subfigure}

    \begin{subfigure}[b]{0.45\textwidth} 
        \centering
        \includegraphics[width=\textwidth]{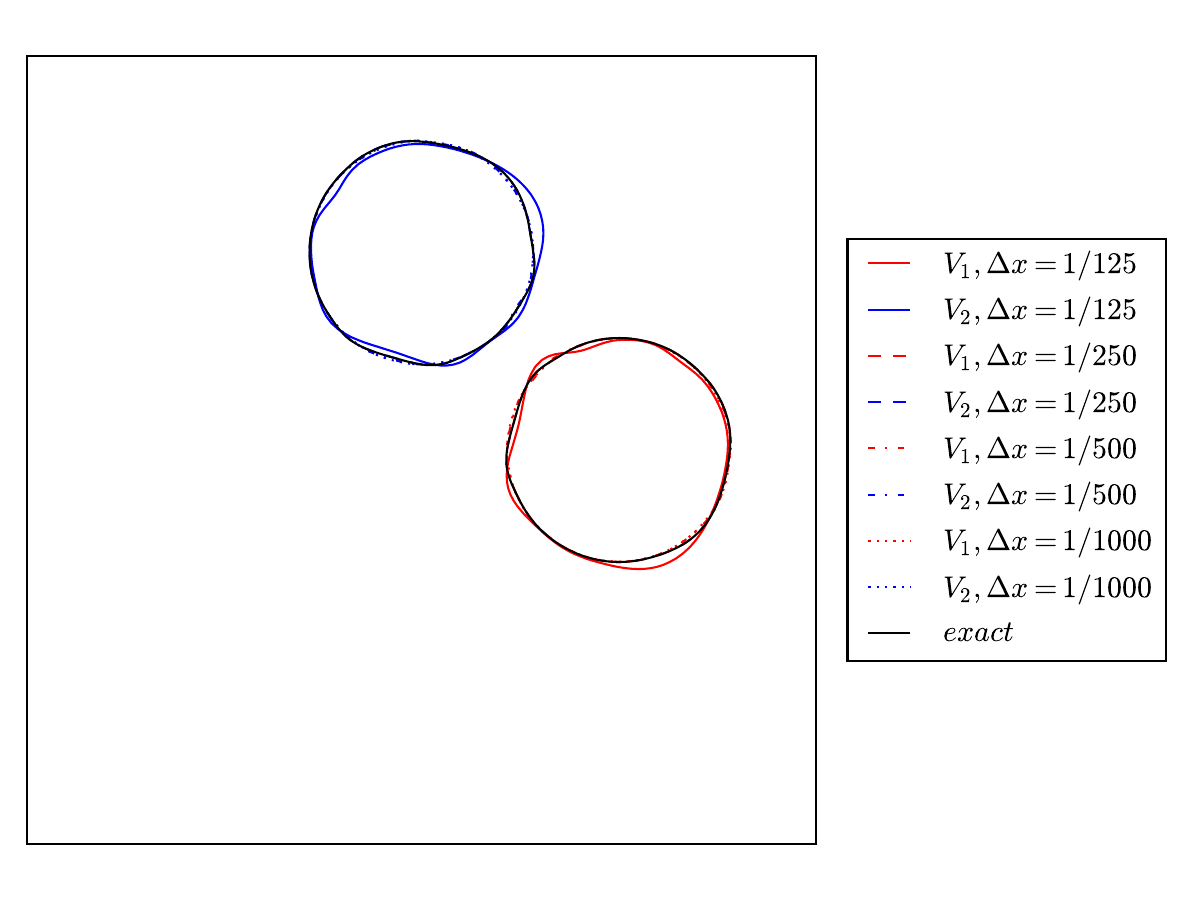}
        \caption{ }
        \label{fig:vortex_end}
    \end{subfigure}
    \begin{subfigure}[b]{0.45\textwidth} 
        \centering
        \includegraphics[width=\textwidth]{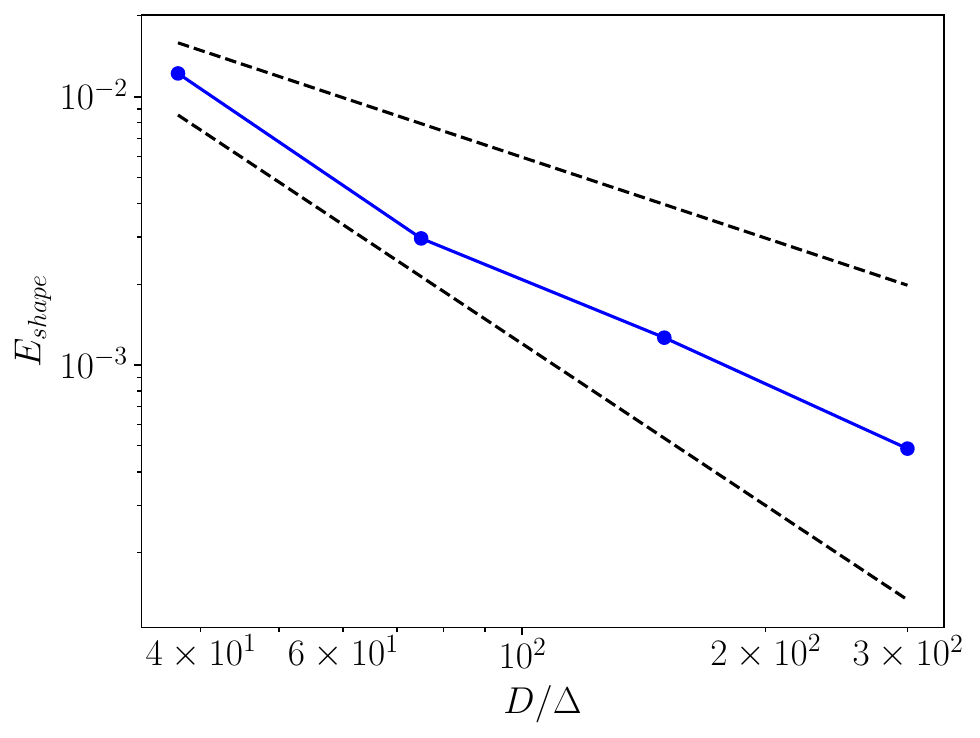}
        \caption{ }
        \label{fig:vortex_error}
    \end{subfigure}
    \caption{(a) Contours of $V=0.5$ at the initial conditions for the droplets in shear flow. (b) Contours of $V=0.5$ for the two droplets in shear flow at $t=2$, the midpoint of the simulation, for various resolutions. (c) Contours of $V=0.5$ for the two droplets in shear flow at $t=4$, the midpoint of the simulation, for various resolutions compared to the exact solution. (d) Plots of the shape error compared to the grid resolution. The top dashed line shows first order convergence while the bottom line shows second order convergence. }
    \label{fig:vortex}
\end{figure}

The simulations were performed on uniform grids in both directions with $N=125$, $250$, $500$, and $1000$ grid points in each direction. The initial droplets are defined with profile:
\begin{align}
V_i = \frac{1}{2}\left(1 + \text{erf}\left( \frac{r - \|\bm{x}-\bm{x}_{c,i} \|^2}{2\Delta}\right)\right),
\end{align}
where $V_i$ is the volume fraction of droplet $i$, $r$ is the radius of the droplets, and $\bm{x}_{c,i}$ is the location of the center of droplet $i$. Figure~\ref{fig:vortex_ic} shows the intial $V_i=0.5$ locations of the droplets. 

\begin{figure}[]
    \centering
    \includegraphics[width=0.5\textwidth,trim={0 0 0 0cm}]{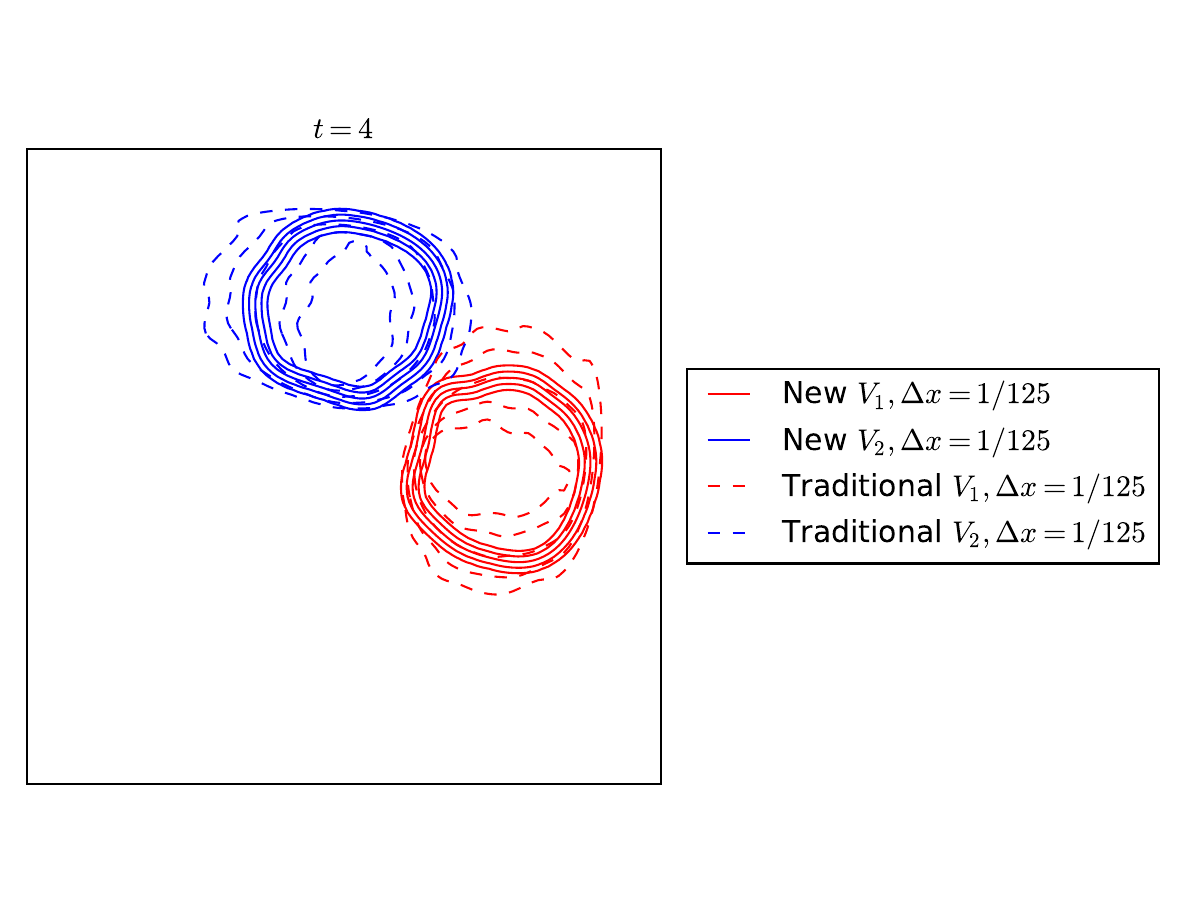}%
    \caption{Comparison of the solutions of the droplets in shear flow at $t=4$ and $\Delta=1/125$ between the solutions with the presented method and the traditional method.}
    \label{fig:vortex_compare}
\end{figure}

Figure~\ref{fig:vortex_mid} shows the shapes of the droplets before the flow reverses. Smaller droplets are seen at the ends of the sheared droplet. As with the previous test cases, the sharpening term creates these droplets. The tail droplets are smaller for the refined cases because their size is based on the grid resolution. The radii are observed to be around $5\Delta$. While sheared, the two droplets remain distinct from one another as as desired. Figure~\ref{fig:vortex_end} shows the final positions of the droplets when they have returned to their initial locations. During the reversed flow, the tail droplets reconnected with the main droplets, showing that the sharpening method does not prevent like materials from joining together. Visibly, only the coarsest resolution shows significant shape deformation. To analyze the shape error of the different resolutions, is computed as: 
\begin{align}
    E_{shape} = \sum_{i=1}^N \sum_{j=1}^N |V_3(i,j,t=T) - V_3(i,j,t=0)|\Delta^2.
\end{align}
The volume fraction of the background material is used in the error calculation because it includes the error of both of the droplet materials. Figure~\ref{fig:vortex_error} shows the error as a function of resolution. It shows between first and second order accuracy, similar to what is reported in~\cite{mirjalili2024conservative}. The reduction in accuracy compared to the 10th order convergence rate of the compact finite difference method is because $\epsilon$ in the sharpening term is proportional to $\Delta$. Figure~\ref{fig:vortex_compare} shows the final solution for the $N=125$ case compared for the presented method compared to the traditional method. The interfaces of the traditional method are much more smeared, which demonstrates that the sharpening term maintains a fixed interface thickness throughout the simulation. This three-material test case was important to ensure that the artificial species diffusive fluxes did not cause density diffusion in the presence of materials with equal densities. As desired, the artificial diffusivity is effectively able to maintain sharp interfaces throughout the shearing of the droplet without unphysical density diffusion.

\subsection{Experimental Comparison: Nitrogen Bubble Collapse in Water}
Finally, we compare results obtained with the presented model to experimental recordings of the shock-induced dynamics of a spherical nitrogen bubble in water~\cite{bokman2025shock}. 
In brief, a microfluidic device generates spherical nitrogen bubbles in a water-filled test chamber. 
A flat copper foil is driven to electrical explosion $6\,\text{mm}$ away from the bubble with a high energy pulsed power driver, which induces a planar shock wave~\cite{strucka2023synchrotron}.
The shock wave propagation and bubble dynamics resulting from their interaction are recorded through in-situ ultra-high-speed synchrotron-based X-ray phase contrast imaging at $5\,\text{Mfps}$.
The peak pressure of the shock can be deduced from its observable speed and the Rankine-Hugoniot jump relation and its temporal decay is estimated through hydrodynamic numerical simulations~\cite{chittenden2004equilibrium}. 
More information on these experimental recordings are available at~\cite{bokman2025shock}. 

\begin{figure}[]
    \centering
    \includegraphics[width=0.5\textwidth,trim={0 0 0 0cm}]{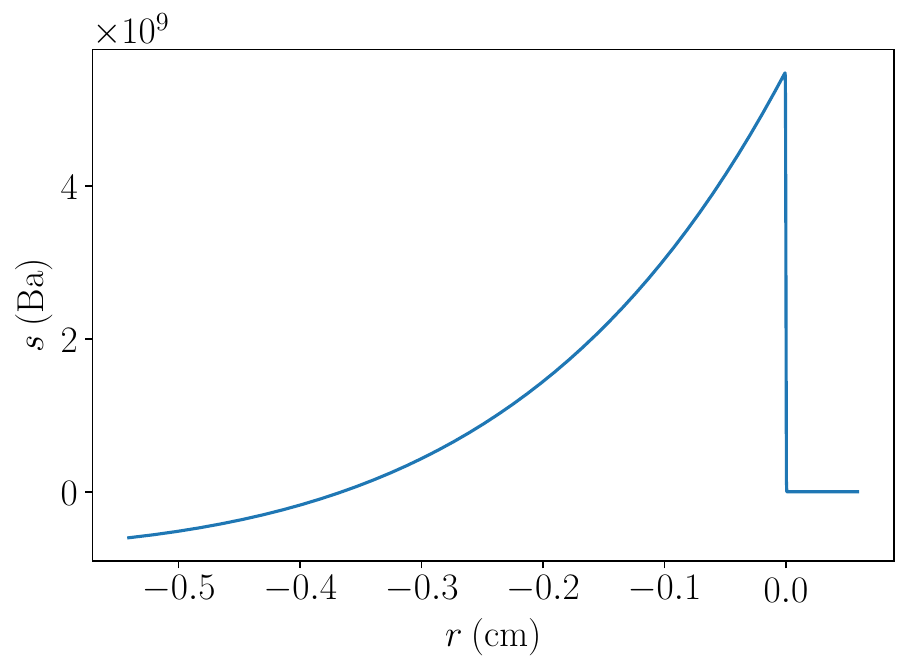}%
    \caption{Plot of the Friedlander pressure pulse fitted to the numerical data from~\cite{bokman2025shock}.}
    \label{fig:ethz_pulse}
\end{figure}

The computational setup uses a 2D axisymmetric domain of size  $r \in [0,0.6]\,\text{cm}$ and $z \in [0,1.2]\,\text{cm}$ with symmetry boundary conditions at $r=0\,\text{cm}$ and $z=0\,\text{cm}$ and outflow boundary conditions at $r=0.6\,\text{cm}$ and $z=1.2\,\text{cm}$. 
A nitrogen-filled half-circle bubble with radius $r=0.047\,\text{cm}$ is centered at $(0,0.6)\,\text{cm}$, which represents the spherical bubble in the axisymmetric coordinates. 
The nitrogen is considered an ideal gas with $M=28.96\,\text{g}/\text{mol}$ and $\gamma=1.4$. 
The water is considered a stiffened gas with the equation of state:
\begin{align}
    p = \rho R T - p_\infty,
\end{align}
where $p_\infty = 7.22\times 10^9\,\text{Ba}$, $\gamma = 2.955$, and $M = 3.368\,\text{g}/\text{mol}$. 
Ambient temperature and pressure are $T_0 = 293.096\,\text{K}$ and $p_0 = 1.01324 \times 10^6\,\text{Ba}$. 
The initial condition pressure, temperature, and velocity are defined for the bubble to be initially at equilibrium at ambient temperature and pressure and the pressure pulse initialized in the domain according to:
\begin{align}
p(r,z,t=0) &= \max(s(z-z_0) + p_0, 0.2 p_0), \\
u(r,z,t=0) &= 0, \\
v(r,z,t=0) &= \frac{s(z-z_0)}{c_s \rho}, \\
T(r,z,t=0) &= T_0 + \frac{s(z-z_0)}{\rho R},
\end{align} 
where $z_0 = 0.541\,\text{cm}$ and $s(z)$ is the shock waveform. The maximum is taken in the definition of the pressure to avoid negative pressures, which lead to instabilities in mixed points. The shock is defined as planar and its exponential decay is approximated by a Friedlander waveform~\cite{dewey2018friedlander}:
\begin{align}
    s(z) = p_b\frac{1}{2}\left( 1- \text{erf}\left(\frac{z}{\delta}\right)\right) \exp\left(d \frac{z}{w}\right)\left(1+\frac{z}{w}\right), \label{eqn:press_prof}
\end{align}
where $z$ is the axial position in the domain, $p_b$ is the shock peak pressure, $\delta$ is the thickness of the front of the pulse, $d$ is a parameter for the decay rate, and $w$ is the spatial width of the pulse. 
For the data in this work, these parameters are $p_b = 5.5\times 10^9\,\text{Ba}$, $\delta = \Delta$, $c = 1$, and $w = 0.3655\,\text{cm}$. 
Figure~\ref{fig:ethz_pulse} shows the pressure pulse.

The artificial diffusivity coefficients used are $C_\mu = 10^{-2}$, $C_\beta = 1$, $C_\kappa = 5.0\times 10^{-2}$, $C_{D,Y}=C_{D,V} = 1.2$,  $\epsilon = 1.5\Delta$, and $\Gamma = 0.5 \max(|\bm{u}|)$. These coefficients were found to be the minimum robust values to stabilize the pressure pulse's impact with the bubble. The simulation was run to a final time of $t=1.2\times 10^{-5}\,\text{s}$. The mesh is a uniform grid of $1920 \times 3840$ points, which corresponds to $300$ grid points per diameter.

\begin{figure}[]
    \centering
    \includegraphics[width=0.5\textwidth,trim={0 0 0 0cm}]{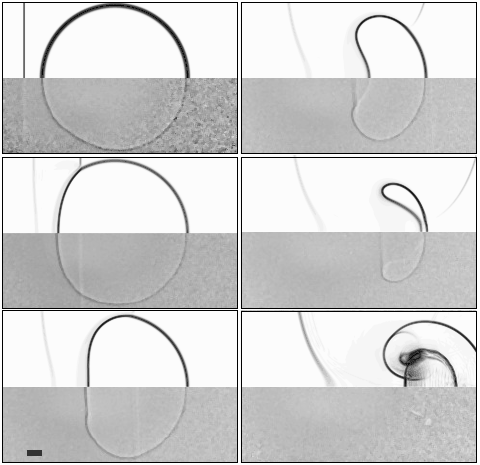}%
    \caption{Comparisons between numerical Schlieren computational images (top) and X-ray phase contrast experimental images (bottom)~\cite{bokman2025shock} throughout the bubble collapse. Results are plotted at times $t=-0.06,0.14,0.34,0.54,0.74,0.94\,\mu \text{s}$. $t$ is defined such that $t=0\,\mu \text{s}$ is when the shock impacts the bubble. The scale bar in the bottom left corner indicates $100\,\mu \text{m}$.}
    \label{fig:num_schlier}
\end{figure}

\begin{figure}[]
    \centering
    \includegraphics[width=0.75\textwidth,trim={0 0 0 0cm}]{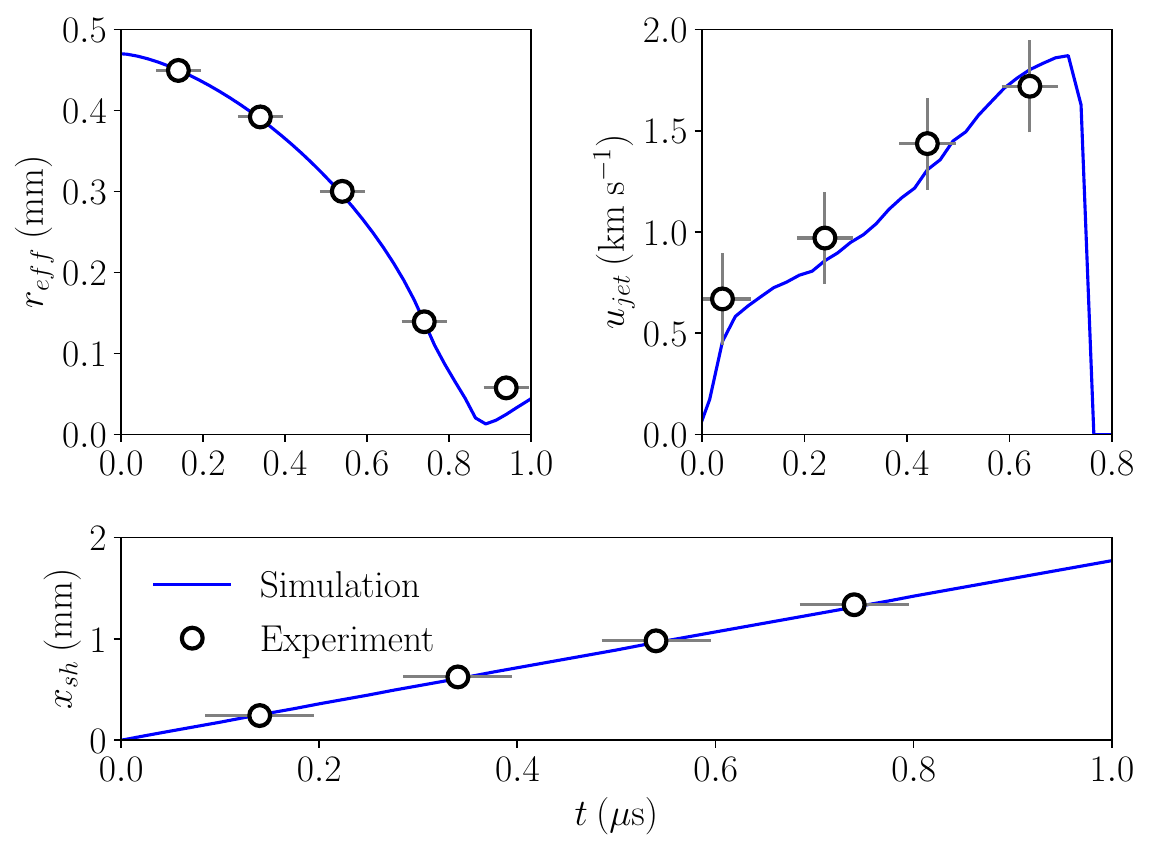}%
    \caption{Comparison between numerical and experimental data for the effective radius of the bubble, $r_{eff}$, jet velocity, $u_{jet}$, and the shock position, $x_{sh}$. $t$ is defined relative to the moment the shock impacts the bubble.}
    \label{fig:ethz_comparison}
\end{figure}

Figure~\ref{fig:num_schlier} shows a comparison between the numerical Schlieren from the simulation with the X-ray phase contrast images from the experiment, yielding good agreement throughout the collapse. The agreement can be further validated by quantifying the bubble compression, jet evolution, and location of the shock wave. The effective radius of the bubble is used to characterize the compression of the bubble. It is defined as:
\begin{align}
    r_{eff} = \sqrt{\frac{A_{\text{nitrogen}}}{\pi}}, 
\end{align}
where $A_{\text{nitrogen}}$ is the area of the visible nitrogen in the $2D$ image. For the experiment, $A_{\text{nitrogen}}$ is computed as the total area of nitrogen in the X-ray phase contrast images. For the simulation, the area is $A_{\text{nitrogen}} = 2 \sum_\Omega V_{\text{nitrogen}}$. The top left plot of Fig.~\ref{fig:ethz_comparison} shows the simulated effective radius compared to the experiment. Good agreement is observed between the experimental data and the simulation during the entire collapse of the bubble. There is some discrepancy between the effective radius after the bubble reinflates around $t=0.9\,\mu \text{s}$ after impact. This is because the four equation model used in this work lacks the expansion and compression of the individual materials that is possible with multi-material models that use additional equations~\cite{SCHMIDMAYER2020109080}. The top-right plot of Fig.~\ref{fig:ethz_comparison} shows a comparison of the velocity of the jet. This comparison shows that the numerical method is able to effectively capture the dynamics of the liquid jet that propagates within the bubble. 
The bottom figure of Fig.~\ref{fig:ethz_comparison} shows the positions of the shock in time of the simulation compared to the experimental data. The agreement confirms that although the initial peak pressure used in this work is 17\% lower than the one used in~\cite{bokman2025shock}, its propagation agrees well with the experimental data. Specifically, the value of peak pressure results in the correct particle velocity of the shock.
Overall, the numerical simulation agrees well with the experiment during the collapse of the bubble, but displays slight discrepancies during its reinflation due to the instantaneous pressure and temperature equilibration of two materials. 

\section{Conclusions}
A novel localized-artificial diffusivity method for the compact finite difference method was presented to better simulate high density ratio and immiscible interfaces. Traditional methods struggle for large density ratios because the diffusivities and diffusive fluxes are targeted at ringing and gradients in the mass fraction, but as the density ratio increases, the mass fraction is further from the species mass, which is the quantity actually being transported. To address this issue, the artificial diffusivity and species diffusion flux were modified to focus on species mass. Additionally, artificial bulk density diffusion was used to regularize the interface to preserve pressure/temperature equilibrium. Furthermore, a sharpening term was added to maintain a finite interface thickness, so that immiscible interfaces can be modeled. These beneficial properties were demonstrated on a number of test problems, which show that the presented method can stably simulate interfaces with density ratios of up to $10^6$. Additionally, it was shown that the sharpening term maintains the interface thickness in flows with shearing and shocks. Finally, the method was compared with experimental data of a Nitrogen bubble collapse in water. Good agreement is seen in capturing the collapse of the bubble. The presented method underpredicted the reinflation of the bubble, which is consistent with literature stating that this is a weakness of the four-equation model. Future work will attempt to improve the multi-material method for the compact finite difference method to address this weakness.

\section{Acknowledgements}
This work was performed under the auspices of the U.S. Department of Energy by Lawrence Livermore National Laboratory under Contract No. DEAC52-07NA27344. This article has been assigned an LLNL document release number (LLNL-JRNL-2003386-DRAFT). The authors are grateful to Shahab Mirjalili and Zoe Barbeau for helpful discussions.

\appendix

\section{Diffuse Interface Offset Between Mass and Volume Fractions}\label{app:offset}
In the diffuse interface method, there are mixed regions between the materials. For a two-material, 1-D interface, we define the location of the mass fraction interface, $x_{c,Y}$, as the location where $Y_l=Y_h=0.5$. The two materials are denoted as $h$ for the heavy material and $l$ for the light material. Similarly, the location of the volume fraction interface, $x_{c,V}$, is the location where $V_l=V_h=0.5$. We will derive the distance between $x_{c,Y}$ and $x_{c,V}$. Let $R=\rho_h/\rho_l$ be the density ratio..
\begin{align}
    \rho_i V_i &= \rho Y_i, \\
    Y_l + Y_h &= 1, \\
    V_l + V_h &= 1, \\
    \rho &= \rho_l V_l + \rho_h V_h, 
\end{align} 
Beginning with the relationship between mass and volume fraction, we can derive this relationship as a function of density ratio:
\begin{align}
    \rho_l V_l &= \rho Y_l, \\
    V_l &= \frac{\rho Y_l}{\rho_l} \\
        &= \frac{(\rho_l V_l + \rho_h V_h)Y_l}{\rho_l} \\
        &= (V_l + R V_h)Y_l \\
        &= (V_l + R(1-V_l))Y_l\\
    V_l + (R-1)V_l Y_l &= R Y_l \\
    V_l &= \frac{R Y_l}{1-Y_l + RY_l}.
\end{align}
Hence, at $x_{c,Y}$, 
\begin{align}
    V_l = \frac{0.5 R}{1-0.5 + 0.5R} = \frac{R}{1+R}. \label{eqn:Yl05}
\end{align}
Let us assume that the volume fraction interface is a hyperbolic tangent in shape centered around $x_{c,V}$ with width, $w$:
\begin{align}
    V_l = \frac{1}{2}\left(1-\tanh\left(\frac{x-x_{c,V}}{w}\right)\right). \label{eqn:Vltanh}
\end{align}
Combining Eq.~\eqref{eqn:Yl05} and Eq.~\eqref{eqn:Vltanh} gives the distance between $x_{c,Y}$ and $x_{c,V}$, which can also be nondimensionalized by the width:
\begin{align}
    \frac{R}{1+R} &= \frac{1}{2}\left(1-\tanh\left(\frac{x_{c,Y}-x_{c,V}}{w}\right)\right), \\
    x_{c,Y} - x_{c,V} &= w \tanh^{-1}\left(-\frac{R-1}{R+1}\right) ,\\
    x_{c,Y} - x_{c,V} &= w \tanh^{-1}(At), \\
    \frac{x_{c,Y} - x_{c,V}}{w} &= \tanh^{-1}(At), \\
\end{align}
where $At=\frac{1-R}{1+R}$ is the Atwood number. A visualization of this offset for various density ratios is shown in Fig.~\ref{fig:offset}

\bibliographystyle{unsrtnat}
\bibliography{references}

\end{document}